\documentclass[longbibliography, showpacs, aps, oneside, twocolumn, prd, amsmath, amssymb, nofootinbib, superscriptaddress]{revtex4-1}
\usepackage{cases}
\usepackage{amsmath}
\usepackage{amssymb}
\usepackage{amsfonts}
\usepackage{dcolumn}
\usepackage{comment}
\usepackage{bm}
\usepackage{bbm}
\usepackage{graphicx}
\usepackage{color}
\usepackage{xcolor}
\usepackage{array}
\usepackage{subfigure}
\usepackage{hyperref}
\usepackage{mathrsfs}
\usepackage{float}

\hypersetup{colorlinks=true,
	breaklinks=true,
	pdfstartview=Fit,
	linkcolor=blue,
	citecolor=blue,
	urlcolor=blue}
\begin{document}

\title{Evading the CMB $\mu$-distortion bound on Supermassive Primordial Black Hole seeds with Non-Gaussian tails }

\author{Sanket Dave}
\email{sd1997@ustc.edu.cn}
\affiliation{Department of Astronomy, School of Physical Sciences, University of Science and Technology of China, Hefei, Anhui 230026, China}
\affiliation{CAS Key Laboratory for Researches in Galaxies and Cosmology, School of Astronomy and Space Science, University of Science and Technology of China, Hefei, Anhui 230026, China}

\author{Sheng-Feng Yan}
\email{sfyan@cdut.edu.cn}
\affiliation{College of Physics, Chengdu University of Technology, Chengdu 610059, China}

\author{Amara Ilyas}
\email{aarks@ustc.edu.cn}
\affiliation{Department of Astronomy, School of Physical Sciences, University of Science and Technology of China, Hefei, Anhui 230026, China}
\affiliation{CAS Key Laboratory for Researches in Galaxies and Cosmology, School of Astronomy and Space Science, University of Science and Technology of China, Hefei, Anhui 230026, China}

\author{Yi-Fu Cai}
\email{yifucai@ustc.edu.cn}
\affiliation{Department of Astronomy, School of Physical Sciences, University of Science and Technology of China, Hefei, Anhui 230026, China}
\affiliation{CAS Key Laboratory for Researches in Galaxies and Cosmology, School of Astronomy and Space Science, University of Science and Technology of China, Hefei, Anhui 230026, China}

\begin{abstract} 
Supermassive black holes (SMBHs) powering quasars at $z \gtrsim 6$ are difficult to grow from stellar mass remnants, motivating seeds from primordial black holes (PBHs) with masses $10^5-10^7 M_{\odot}$. This range is constrained by the COBE/FIRAS bound on the CMB $\mu$-distortion, which limits the small-scale curvature variance to $\sigma_\zeta^2 \lesssim 10^{-4}$. For Gaussian perturbations, the variance fixes the far tail of the one-point probability distribution function (PDF), making the PBH abundance negligible. We call this the Gaussian barrier. The barrier can be evaded only if the variance probed by the distortion is decoupled from the tail probability controlling collapse. We implement this idea in the non-perturbative $\delta N$ formalism and relate asymptotic PDF tails to the global shape of the $\delta N$ map. Four Gaussian-cored families are analyzed: generalized-normal, stretched-exponential, power-law, and log-normal tails. After standardizing each family to unit variance, we impose the FIRAS cap and compute the distortion-limited PBH abundance in the tail-shape parameter space. The ordinary exponential tail produced by standard single-field non-attractor dynamics is still too light to reopen the seed window. Algebraic tails from fractional-potential dynamics, and sufficiently heavy log-normal tails treated as a phenomenological proxy for multiplicative dynamics, can supply seed-relevant abundances while respecting the distortion bound.
\end{abstract}

\maketitle

\section{Introduction}
\label{sec:intro}

Luminous quasars at $z \gtrsim 6$ show that black holes of mass $\sim 10^8-10^{10} M_{\odot}$ existed within the first billion years~\cite{Mortlock2011,Banados2018,Wang2021,Yang2020,Fan2023}. More than one hundred such quasars are now known, with the current highest-redshift example at $z \simeq 7.64$~\cite{Wang2021}, and the Event Horizon Telescope image of the SMBH in M87 has confirmed that such objects are generic to massive galaxies~\cite{EHT2019M87,Volonteri2021}. Explaining their early assembly remains a central problem in galaxy formation and cosmology~\cite{Volonteri2010,InayoshiVisbalHaiman2020}.

The difficulty is one of timing. Under Eddington-limited accretion the mass grows as $M(t)=M_{\rm seed}\,e^{t/t_s}$, with a Salpeter time
\begin{equation*}
t_s\simeq\frac{\epsilon}{1-\epsilon}\,\frac{\sigma_T\,c}{4\pi\,G\,m_{\rm pr}}\approx 50\,\mathrm{Myr}
\end{equation*}
for radiative efficiency $\epsilon\simeq0.1$, Thomson cross section $\sigma_T$, proton mass $m_{\rm pr}$; the Salpeter time is independent of the black hole mass~\cite{Maiolino2024}. Growing a $\sim 10^2 M_{\odot}$ remnant to $10^{10} M_{\odot}$ requires about $0.9 ~\mathrm{Gyr}$, comparable to the cosmic age at $z \simeq 6-7$. Because the first stars form only $\sim 0.1-0.2~ \mathrm{Gyr}$ after the Big Bang, light seeds require either sustained super-Eddington growth or unusually favorable environments~\cite{Volonteri2010,InayoshiVisbalHaiman2020}. Heavy-seed channels have been proposed, including population III remnants~\cite{MadauRees2001}, direct collapse of pristine gas into $10^{4} - 10^{6}\,M_\odot$ objects~\cite{BrommLoeb2003,BegelmanVolonteriRees2006,LodatoNatarajan2006}, and runaway collisions in dense clusters~\cite{PortegiesZwartMcMillan2002}, but each requires fine-tuned conditions. The corresponding intermediate-mass black holes have not been securely identified in the local Universe~\cite{GreeneStraderHo2020}. JWST detections of accreting black holes at $z \gtrsim 8$ and the abundant ``little red dots'' (LRDs) have sharpened this issue, with inferred masses favoring seeds in the $10^5-10^7 \, M_{\odot}$ range~\cite{Larson2023,Kocevski2023,Matthee2024,Greene2024,Maiolino2024,Bogdan2024}. Primordial-seed pathways to these objects face the same distortion obstruction under Gaussian statistics~\cite{DeLuca:2025nao}.

However, the timing problem could be eased if the black hole seeds formed very early in the universe. PBHs form when rare curvature perturbations collapse after horizon re-entry~\cite{Zeldovich1967,Hawking1971,CarrHawking1974,Carr1975}. PBHs have been studied for decades as a dark-matter candidate and as a probe of small-scale inflationary physics~\cite{Khlopov2010,Sasaki2018review,CarrKuhnel2020,GreenKavanagh2021,Villanueva2021,CarrKohriSendoudaYokoyama2021,Carr2026}, and a seed-mass population of $M_{\mathrm{PBH}} \sim 10^5- 10^7 M_{\odot}$ corresponds to modes with $k \sim 10^2-10^4 \mathrm{Mpc}^{-1}$ during radiation domination. Such a population is sufficient to alleviate the timing problem above. Generating them requires a localized enhancement of small-scale primordial fluctuations, as can occur in many early-universe mechanisms~\cite{IvanovNaselskyNovikov1994,GarciaBellidoRuizMorales2017,MotohashiHu2017,Kannike2017,BallesterosTaoso2018,germani2017primordialblackholesinflection,Mishra_2020,Hertzberg_2018,Cicoli_2018,OzsoyTasinato2023,Cai:2018tuh,Cai:2020ovp,Fu:2022ypp,Yi:2020cut,Di:2017ndc,Cai:2023uhc,Chen:2022fda,Pi:2022zxs,Pi2024NonGaussianitiesReview}. Such scenarios also source a stochastic gravitational-wave background at second order~\cite{SaitoYokoyama2009,KohriTerada2018,Domenech2021,Cai:2019jah,Baumann:2007zm,Yuan:2021qgz,Chen:2019xse,Cai:2019bmk}; nonlinear dynamics in low-scale inflation can likewise source gravitational waves and act as a PBH formation channel~\cite{Masubuchi:2026eau}.

The same scales are constrained by CMB spectral distortions. Photon diffusion damps acoustic waves and injects energy into the photon bath; in the $\mu$ era this produces a chemical-potential distortion~\cite{SunyaevZeldovich1970,Daly1991,HuScottSilk1994,ChlubaSunyaev2012,KhatriSunyaevChluba2012,ChlubaErickcekBenDayan2012,Jeong2014,Chluba2021Review}. The COBE/FIRAS limit $\mu<9 \times 10^{-5}$ therefore bounds the curvature variance on the seed-relevant scales~\cite{Fixsen1996,Nakama2018,KohriNakamaSuyama2014,Bianchini2022muDistortion}. For Gaussian statistics this variance also fixes the far tail, driving the PBH abundance far below the level needed for seeds. We call this obstruction the Gaussian barrier. Non-Gaussian statistics, however, can evade it.

In this work we revisit the barrier from the standpoint of the full one-point statistics of the curvature perturbation. The $\mu$-distortion is, at leading order in the acoustic-damping calculation, controlled by the power spectrum and hence by a variance of the distribution of the primordial perturbations. PBH formation, in contrast, is controlled by the probability that a smoothed perturbation exceeds the collapse threshold, i.e.\ on the deep tail of the PDF. For Gaussian fluctuations, the variance fixes the tail; while for non-Gaussian statistics, the two are decoupled. The relevant tail lies $50-100$ standard deviations away from the core, where a perturbative expansion in $f_{\rm NL}$ and higher cumulants is no longer reliable~\cite{Celoria2021,Hooshangi2022,DaviesCarrilhoMulryne2022,FerranteFranciolini2023}. We therefore work with non-perturbative PDF tails generated or motivated by the $\delta N$ map, standardize each family to fixed total variance, impose the FIRAS cap, and ask which tail shapes can still yield a seed-relevant PBH abundance. This connects the tail taxonomy of Refs.~\cite{Hooshangi2022,Hooshangi2023} with the spectral-distortion analyses of Refs.~\cite{Nakama2018, UnalKovetzPatil2021,Byrnes:2024vjt,Sharma:2024img,Pritchard:2025yda,Hooper2024} in a single variance-capped calculation.

We proceed in the following steps. First, we write the calculation as a variance cap plus a standardized tail integral; in this form the Gaussian suppression follows from the statistics alone, independently of the model. Second, we use the non-perturbative $\delta N$ formalism to organize the admissible asymptotic tails, and study four Gaussian-cored representatives: generalized-normal, stretched-exponential, power-law, and log-normal~\cite{StarobinskyDeltaN1985,SasakiStewart1996,WandsMalikLythLiddle2000,LythRodriguez2005}. Third, we translate the resulting $\beta_{\max}(M)$ into viable regions of tail-shape parameter space. Two approximations hold throughout: the fixed-$\zeta_c$ threshold is a simple proxy for a compaction-function treatment, and the log-normal tails do not yet follow from a derived model.

Section~\ref{sec:background} derives the mass–scale relation, the spectral distortion bound, and the standardized tail integral. Section~\ref{sec:PDF_analyses} develops the $\delta N$ tail dictionary and the four tail families. Section~\ref{sec:Results} presents the distortion-capped abundances and viable parameter space. Section~\ref{sec:conclusion} discusses the implications and open questions. We work in units with $c=\hbar=1$ and quote masses in solar units.
\section{Spectral-distortion bound and the Gaussian barrier}
\label{sec:background}
 
Cosmic inflation~\cite{Starobinsky_1980_Inflation,Guth_1981_Inflation,Linde_1982_New_Inflation,Steinhardt_1982_New_Inflation,Linde_1983_Chaotic_Inflation,Linde_1990_Inflation_Book} provides the standard account of the origin of primordial perturbations, generating a nearly Gaussian, nearly scale-invariant spectrum of curvature fluctuations from the stretching of quantum vacuum modes~\cite{Mukhanov_1981_Q,Hawking_1982_Q,Starobinsky_1982_Q,Guth_1982_Q,MukhanovFeldmanBrandenberger1992}. On the scales probed by the CMB and BAO the power spectrum is
\begin{equation}
\mathcal{P}_\zeta(k)=A_s\left(\frac{k}{k_*}\right)^{n_s-1},
\end{equation}
with $A_s\approx2.1\times10^{-9}$ at $k_*=0.05\,\mathrm{Mpc}^{-1}$. Planck gives $n_s=0.9649\pm0.0042$~\cite{Planck:2018vyg,Planck_2018_Inflation}, rising to $n_s=0.9728\pm0.0029$ when ACT DR6, SPT, BK and DESI BAO data are combined with Planck~\cite{ACT_DR6_Louis2025, Balkenhol:2025wms, BICEP_Keck}. These measurements probe only a limited range of scales; much smaller scales are observationally unconstrained, leaving room for a localized enhancement of power spectrum resulting in PBH formation.

During radiation domination a PBH forms when an overdense region re-enters the horizon with a smoothed amplitude above a critical threshold, and its mass is a fraction $\gamma\simeq0.2$ of the horizon mass at re-entry~\cite{Carr1975,Sasaki2018review}. This fixes the relation between the PBH mass and the comoving scale,
\begin{equation}
\begin{split}
M_{\rm PBH}(k)\simeq{}&52\,M_\odot\left(\frac{\gamma}{0.2}\right)\left(\frac{g_*}{10.75}\right)^{-1/6}\\
&\times\left(\frac{k}{2.9\times10^{5}\,{\rm Mpc}^{-1}}\right)^{-2},
\end{split}
\label{Eq:mass_k_relation}
\end{equation}
so that the seed range $M_{\rm PBH}\sim10^{5}$--$10^{7}\,M_\odot$ corresponds to $k\sim10^{2}$--$10^{4}\,{\rm Mpc}^{-1}$. These scales are three to five orders of magnitude smaller than what CMB temperature and polarization anisotropies can probe, but they fall within the window in which spectral distortions are most sensitive to the small-scale power. At $z\gtrsim2\times10^{6}$ double-Compton and bremsstrahlung processes maintain a blackbody spectrum; in the window $5\times10^{4}\lesssim z\lesssim2\times10^{6}$ these number-changing processes freeze out while Compton scattering still redistributes energy, so the energy released by the damping of small-scale modes produces a Bose--Einstein spectrum with a chemical potential $\mu$~\cite{SunyaevZeldovich1970,ChlubaSunyaev2012,Nakama2018}.

The distortion is a weighted integral of the curvature power spectrum,
\begin{equation}
\begin{split}
\mu \simeq 2.2 \int &\mathrm{d}\ln k \; \mathcal{P}_\zeta(k) \\
&\times \left[ e^{-k/(5400\,{\rm Mpc}^{-1})} - e^{-(k/(31.6\,{\rm Mpc}^{-1}))^{2}} \right],
\end{split}
\label{Eq:mu_window}
\end{equation}
which for a narrow or sharply peaked spectrum of variance $\sigma_\zeta^{2}$ reduces to~\cite{Nakama2018}
\begin{equation}
\mu(M)\simeq2.2\,\sigma_\zeta^{2}\,W_\mu(M),
\label{eq:mu_narrow}
\end{equation}
with the window function expressed in terms of PBH mass $M$,
\begin{equation}
\begin{split}
W_\mu(M)\equiv
\exp\!\left(-\sqrt{\frac{1.5\times10^{5}\,M_\odot}{M}}\right) \\ 
-\exp\!\left(-\frac{4.5\times10^{9}\,M_\odot}{M}\right).
\label{eq:W_mu}
\end{split}
\end{equation}
We adopt this narrow-feature approximation as our baseline together with the FIRAS limit $\mu<9\times10^{-5}$~\cite{Fixsen1996}; it captures the variance-controlled nature of the bound, and we return to its limitations in Sec.~\ref{sec:conclusion}. 

\begin{figure*}[t]
\centering
\subfigure[ Distortion-capped variance $\sigma_{\zeta,\max}^{2}(M)$.]{\includegraphics[width=3.1in]{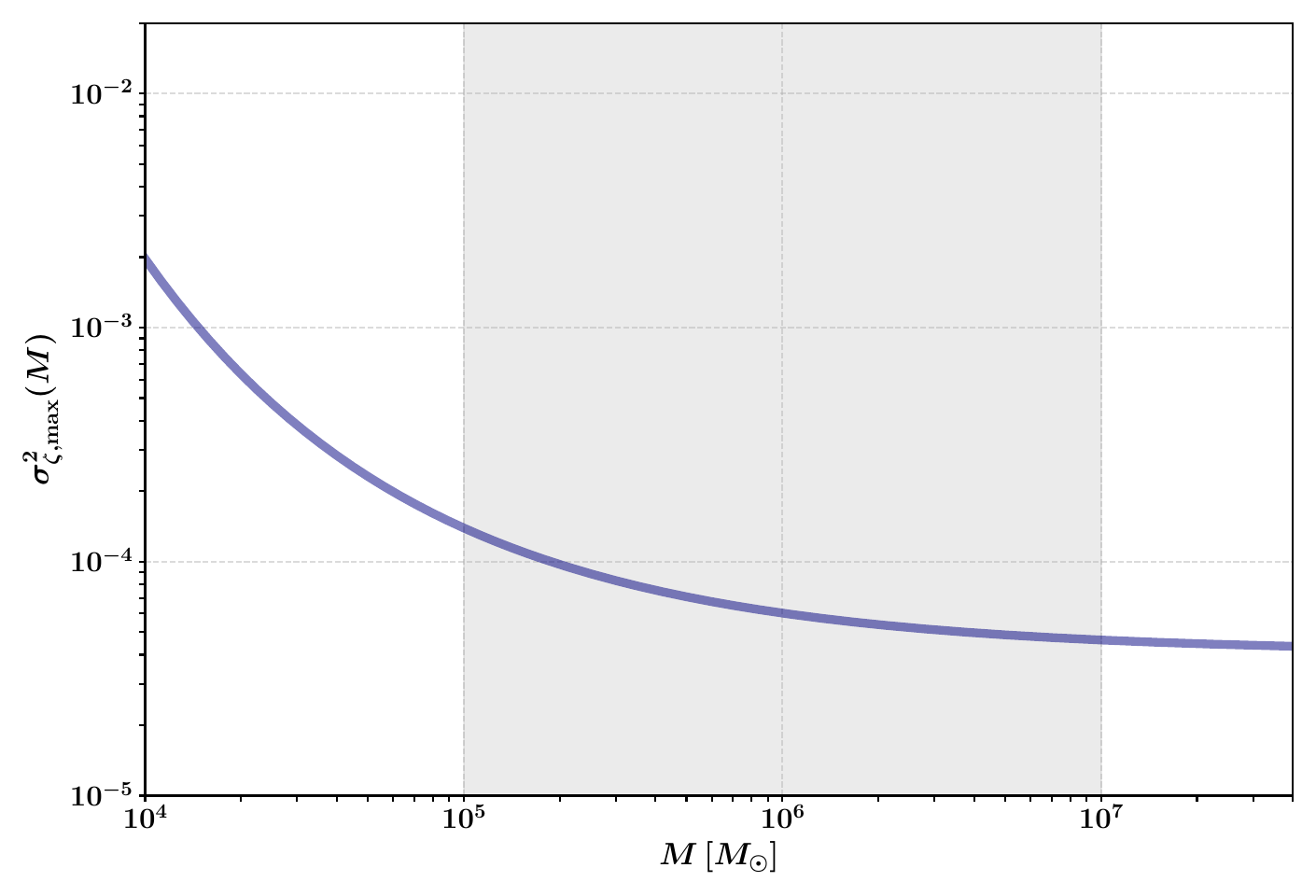}}
\hfill
\subfigure[ Standardized PDFs of the tail families.]{\includegraphics[width=3.1in]{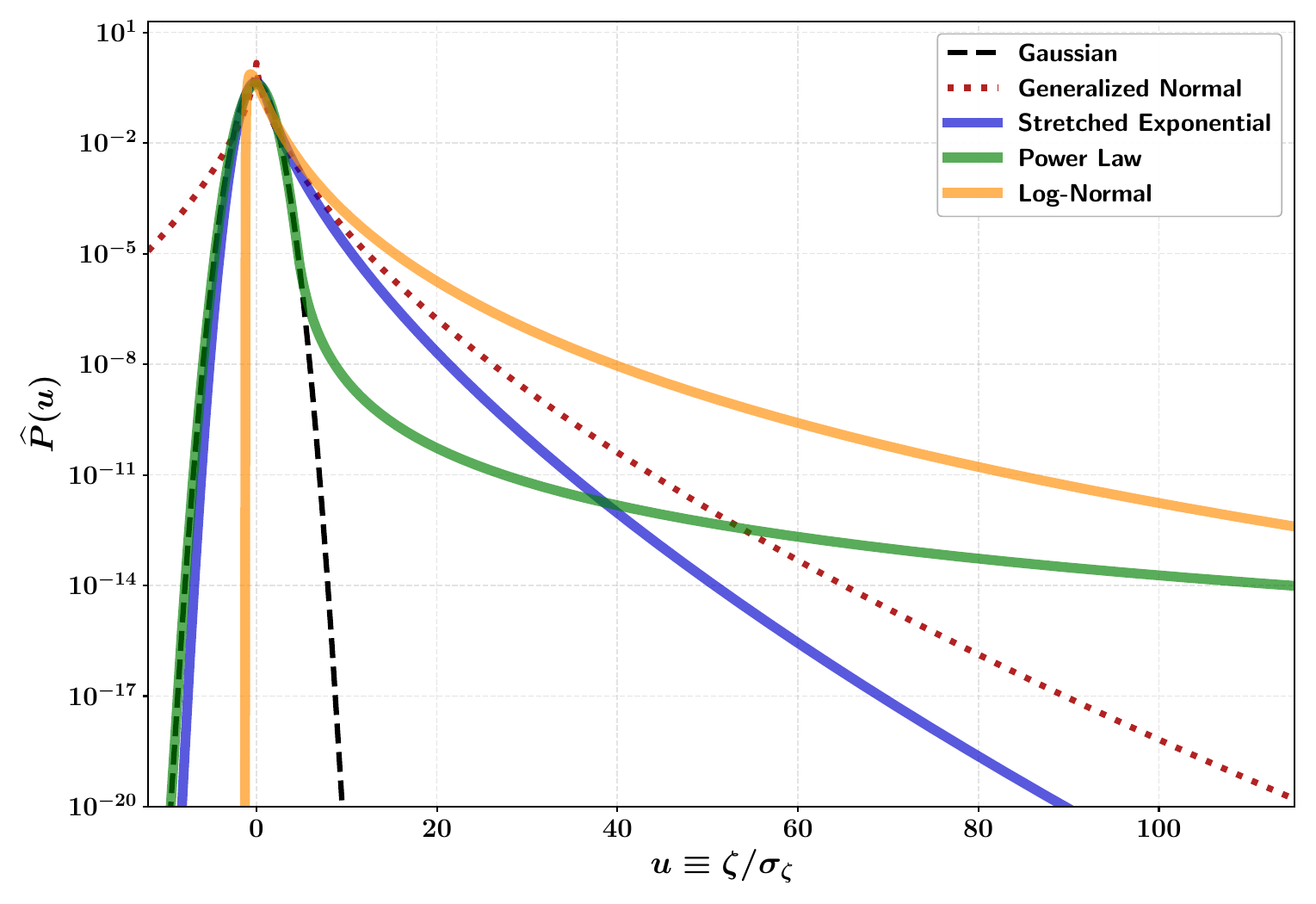}}
\caption{\textbf{(a)} Maximum variance $\sigma_{\zeta,\max}^{2}(M)$ permitted by the FIRAS $\mu$-distortion bound for a narrow power spectrum, Eq.~\eqref{eq:sigma_cap_mu}. The shaded band marks the SMBH-seed range $10^{5}$--$10^{7}\,M_\odot$, where $\sigma_{\zeta,\max}^{2}\lesssim10^{-4}$ and the corresponding standardized threshold is $u_c\sim50$--$100$. \textbf{(b)} Standardized one-point PDFs $\widehat P(u)$ with $u\equiv\zeta/\sigma_\zeta$, for a Gaussian (black dashed) and for one representative of each non-Gaussian family studied here: generalized-normal ($p=0.6$), stretched-exponential ($p=0.6$), power-law ($q=3.5$), and log-normal ($s=0.7$). All curves share the same unit variance, so the panel isolates differences in tail shape: at the large $u$ relevant for collapse the Gaussian is exponentially suppressed while the heavy tails retain support.}
\label{fig:overview}
\end{figure*}

The following issues must be treated carefully. A heavy tail contributes to the total variance, so each family is standardized to unit total variance before the FIRAS cap is applied (Sec.~\ref{subsec:pipeline}). In addition, the conversion from primordial statistics to $\mu$ can receive non-Gaussian corrections beyond the leading acoustic-damping result. Refs.~\cite{UnalKovetzPatil2021,Sharma:2024img,Byrnes:2024vjt,Pritchard:2025yda} find these corrections to be small for mildly non-Gaussian spectra and model dependent only in the strongly non-Gaussian regime. So, we retain the leading-order relation~\eqref{eq:mu_narrow} throughout. This fixes the absolute level of the cap, not the relative standing of the families: every distribution is capped by the same $\mu$, so neglecting the correction can rescale $\sigma_{\zeta,\max}$ by a common factor but cannot change the ordering of tail shapes that is our main result. The bound caps the variance at each mass scale

\begin{equation}
\sigma_{\zeta,\max}^{2}(M)=\frac{\mu_{\rm lim}}{2.2\,W_\mu(M)},
\label{eq:sigma_cap_mu}
\end{equation}
shown in the left panel of Fig.~\ref{fig:overview}. Across the seed window the cap is $\sigma_{\zeta,\max}^{2}\lesssim10^{-4}$, and it tightens with increasing mass.

The cosmological abundance of PBHs at formation is parameterized by the mass fraction $\beta(M)=\rho_{\rm PBH}/\rho_{\rm tot}\big|_{\rm form}$, which in the Press--Schechter approach~\cite{Press_Schechter,YoungByrnes2013,Young2014} is the integral of the one-point PDF above the threshold,
\begin{equation}
\beta(M)\simeq2\int_{\zeta_c}^{\infty}P_\zeta(\zeta)\,\mathrm{d}\zeta,
\label{eq:beta_def}
\end{equation}
where the factor of two is conventional and we adopt the fiducial value $\zeta_c=0.67$~\cite{Nakama2018}. The present-day PBH fraction follows from~\cite{Sasaki2018review}
\begin{equation}
\begin{split}
\beta\simeq1.1\times10^{-8}\,\gamma^{-1/2}
&\left(\frac{g_*}{10.75}\right)^{1/4}\\
&\times\left(\frac{\Omega_{\rm DM}}{0.27}\right)^{-1}
\left(\frac{M}{30\,M_\odot}\right)^{1/2}f_{\rm PBH}.
\end{split}
\label{eq:beta_to_f}
\end{equation}
$f_{\rm PBH}$ is the present-day dark matter fraction in form of PBHs. For a Gaussian distribution,
\begin{equation}
\beta_G(M)=\mathrm{erfc}\!\left(\frac{\zeta_c}{\sqrt{2}\,\sigma_\zeta}\right)
\simeq\sqrt{\frac{2}{\pi}} \frac{\sigma_\zeta}{\zeta_c}
\exp\!\left(-\frac{\zeta_c^{2}}{2\sigma_\zeta^{2}}\right).
\label{Eq:beta_gaussian}
\end{equation}
With $\sigma_\zeta^{2}\lesssim10^{-4}$ from Eq.~\eqref{eq:sigma_cap_mu} and $\zeta_c\sim\mathcal{O}(1)$, the exponent is of order $10^{3}$ and $\beta_G$ is negligible for seeding purposes~\cite{Carr2026}. The seed requirement (quantified in Sec. \ref{subsec:results_limits}) spans $\beta \sim 10^{-19}$ for one seed per high-$z$ quasar up to $\sim 10^{-13}$ for a whole population origin. We adopt $\beta \sim 10^{-15}$ as a conservative reference, and even this lies vastly above the Gaussian value. This is the Gaussian barrier: the variance that sets the distortion also fixes the collapse tail. Evading the barrier requires breaking that link.

To separate the variance from the shape of the tail we work with the standardized variable
\begin{equation}
u\equiv\frac{\zeta}{\sigma_\zeta}.
\label{eq:u_def}
\end{equation}
Probability is invariant under this rescaling, $P_\zeta(\zeta)\,\mathrm{d}\zeta=\widehat P(u)\,\mathrm{d}u$; with $\mathrm{d}\zeta=\sigma_\zeta\,\mathrm{d}u$ this gives
\begin{equation}
P_\zeta(\zeta\,|\,M,\theta)=\frac{1}{\sigma_\zeta(M)}\,\widehat P(u;\theta),
\end{equation}
where $\theta$ collects the shape parameters. The Jacobian factor $\sigma_\zeta^{-1}$ ensures that $\widehat P$ is normalized in $u$, and, since $\langle\zeta\rangle=0$ and $\langle\zeta^{2}\rangle=\sigma_\zeta^{2}$, that it has zero mean and unit variance,
\begin{equation}
\int\widehat P\,\mathrm{d}u=1,\quad
\int u\,\widehat P\,\mathrm{d}u=0,\quad
\int u^{2}\,\widehat P\,\mathrm{d}u=1,
\label{eq:standardized_constraints}
\end{equation}
so that all amplitude (and hence distortion) information is carried by $\sigma_\zeta(M)$ and all non-Gaussian information by $\widehat P(u;\theta)$. Saturating the distortion bound, $\sigma_\zeta(M)\to\sigma_{\zeta,\max}(M)$, the collapse threshold maps to
\begin{equation}
u_c(M)\equiv\frac{\zeta_c}{\sigma_{\zeta,\max}(M)},
\label{eq:uc_def}
\end{equation}
and the distortion-capped abundance is determined by the standardized collapse function $\mathcal{S}(u;\theta)\equiv\int_u^{\infty}\widehat P(v;\theta)\,dv$,
\begin{equation}
\beta_{\max}(M;\theta)=2\,\mathcal{S}\big(u_c(M);\theta\big).
\label{eq:beta_max}
\end{equation}

Equation~\eqref{eq:beta_max} is the central relation of this paper. Across the seed window $u_c(M)$ takes values of $\sim50$--$100$ (Fig.~\ref{fig:overview}, left panel, via Eq.~\eqref{eq:uc_def}), so collapse samples the distribution tens of standard deviations into the tail, and $\beta_{\max}$ is exponentially sensitive to the asymptotic form of $\widehat P$. The right panel of Fig.~\ref{fig:overview} previews the consequence: a Gaussian falls off as $e^{-u^{2}/2}$ and is negligible at such $u$, whereas the heavier families introduced below retain appreciable weight there. The rest of the paper makes this comparison quantitative.

\section{Non-Gaussian tails and the \texorpdfstring{$\delta N$}{dN} formalism}
\label{sec:PDF_analyses}

The $\mu$-distortion bound and the PBH-formation criterion probe disjoint sectors of the primordial one-point statistics. The former is controlled by the second moment (Sec.~\ref{sec:background}), the latter by the collapse probability beyond $\zeta_c$. This section develops the tools needed for that separation. We first argue (Sec.~\ref{subsec:nonpert}) that the relevant tails are intrinsically non-perturbative, so that the perturbative $f_\mathrm{NL}$ hierarchy is an inadequate organizing tool. Then we show (Sec.~\ref{subsec:taxonomy}) that the $\delta N$ formalism maps generic classes of inflationary dynamics to classes of PDF tails. Families A–D are Gaussian-cored representatives of that dictionary.

\subsection{Why the tail is non-perturbative}
\label{subsec:nonpert}

On super-horizon scales the comoving curvature perturbation equals the perturbation in the local number of $e$-folds~\cite{StarobinskyDeltaN1985,SasakiStewart1996,Pi_2023},
\begin{equation}
\zeta(\mathbf{x})=\delta N\big(\delta\phi(\mathbf{x}),\delta\pi(\mathbf{x})\big)\,,
\quad
\delta\phi\sim\mathcal{N}(0,\sigma_{\delta\phi}^2),
\label{eq:deltaN_master}
\end{equation}
where $\delta\phi$ and $\delta\pi\equiv-\delta(\mathrm{d}\phi/\mathrm{d}N)$ are the Gaussian field and velocity fluctuations on the initial flat slice. The non-Gaussianity of $\zeta$ is carried entirely by the (generally nonlinear) map $\delta N$; the underlying field statistics remain Gaussian. Expanding about the mean reproduces the familiar local series~\cite{LythRodriguez2005},
\begin{equation}
\zeta=\zeta_g+\frac{3}{5}f_\mathrm{NL}\,\zeta_g^2+\dots\,,
\quad
\zeta_g\equiv N'\,\delta\phi\,,
\quad
f_\mathrm{NL}=\frac{5}{6}\,\frac{N''}{N'^{2}}\,,
\label{eq:fNL_expansion}
\end{equation}
with primes denoting derivatives of the $e$-fold number with respect to the field value.

For the present problem, this expansion is inadequate. The distortion cap gives $\sigma_\zeta\lesssim10^{-2}$ for seed masses, as shown in Fig.~\ref{fig:overview}(a), so collapse occurs at $u_c\equiv\zeta_c/\sigma_\zeta\sim50\text{--}100$. This is far outside the perturbative core around which Eq.~\eqref{eq:fNL_expansion} is defined. The relevant field excursions may also exceed the radius of convergence of the Taylor series in $\delta\phi$~\cite{Celoria2021,Hooshangi2022}. More importantly, a cumulant expansion changes the tail and the variance together, whereas the present question requires the tail probability to vary at fixed $\sigma_\zeta^2$ (and hence $\mu$)~\cite{Byrnes:2024vjt,Sharma:2024img}. One must therefore work with the exact map~\eqref{eq:deltaN_master}, as emphasized for single-field models in Refs.~\cite{CaiEtAl2022,DaviesCarrilhoMulryne2022,FerranteFranciolini2023}.

It is convenient to characterize a tail by its local logarithmic slope,
\begin{equation}
D(\zeta)\equiv-\frac{\mathrm{d}\ln P_\zeta(\zeta)}{\mathrm{d}\zeta},
\qquad
D_\infty\equiv\lim_{\zeta\to\infty}D(\zeta)\,,
\label{eq:Dinfty}
\end{equation}
which separates the regimes relevant below~\cite{Hooshangi2023}: a Gaussian has $D(\zeta)=\zeta/\sigma_\zeta^2\to\infty$; a pure exponential $P\propto e^{-k\zeta}$ has $D_\infty=k$; and both a power law $P\propto\zeta^{-(q+1)}$ and a stretched exponential $P\propto e^{-c\zeta^p}$ with $0<p<1$ have $D_\infty=0$. We will refer to the latter two as \emph{heavy} (sub-exponential) tails. As shown in Sec.~\ref{sec:Results}, only heavy tails, $D_\infty=0$, reopen the SMBH-seed window under the FIRAS bound.

\subsection{A tail dictionary from \texorpdfstring{$\delta N$}{dN}}
\label{subsec:taxonomy}

Because $\delta\phi$ is Gaussian, the PDF of $\zeta$ follows from Eq.~\eqref{eq:deltaN_master} by a change of variables,
\begin{equation}
P_\zeta(\zeta)=\sum_i\frac{P_G\!\big(\delta\phi_i(\zeta)\big)}{\big|\,\mathrm{d}\zeta/\mathrm{d}\delta\phi\,\big|_{\delta\phi_i}},
\label{eq:change_of_variables}
\end{equation}
the sum running over branches solving $\delta N(\delta\phi_i)=\zeta$. The asymptotic tail is therefore fixed by the \emph{global} shape of the map, not by its Taylor coefficients. A small set of generic map behaviors yields a correspondingly small set of tail shapes, summarized in Table~\ref{tab:dictionary}: a linear map returns a Gaussian; a logarithmic (saturating) map, characteristic of ultra-slow-roll and upward-step transitions, returns an exponential tail; a non-attractor quadratic map returns a (lighter) Gumbel/double-exponential tail; a map with an integrable singularity at a finite field value returns a power law; and a multiplicative map returns a log-normal. Families A--D below are the standardized, Gaussian-cored representatives of the heavy entries of this dictionary; the Gumbel entry, which is lighter than Gaussian on the collapse side and therefore irrelevant for seeding, is listed for completeness only and is not used in our abundance analysis. The Gaussian core keeps the typical, perturbative fluctuations consistent with CMB-scale measurements, since only rare fluctuations probe the modified branch.
\begin{table}[t]
\centering
\renewcommand{\arraystretch}{1.3}
\caption{Dictionary between the asymptotic shape of the $\delta N$ map (at large $\delta\phi$), the resulting tail of the curvature-perturbation PDF, and its asymptotic log-slope $D_\infty$ [Eq.~\eqref{eq:Dinfty}]. Only the sub-exponential, power-law and log-normal tails in our analysis, where $D_\infty=0$, reopen the seed window. The exponential and Gumbel entries follow from the logarithmic duality of Refs.~\cite{Pi_2023,InuiEtAl2024}; the power-law entry from Refs.~\cite{Hooshangi2022,Hooshangi2023}.}
\label{tab:dictionary}
{\setlength{\tabcolsep}{4pt}
\begin{tabular}{|c|c|c|}
\hline
\textbf{ \boldsymbol{$\delta N$} map (large $\delta\phi$)} & \textbf{Tail of $P_\zeta$} & \boldsymbol{$D_\infty$} \\
\hline
Linear & Gaussian & $\infty$ \\
\hline
Logarithmic / saturating & \shortstack{$e^{-k\zeta}$ \\(exponential)} & $k$ \\
\hline
Non-attractor quadratic & \shortstack{$e^{-c\,e^{k\zeta}}$ \\ (Gumbel) }& $\infty$ \\
\hline
Singular at finite $\delta\phi_{\rm max}$ & \shortstack{$\zeta^{-(q+1)}$ \\ (power law)} & $0$ \\
\hline
Multiplicative & \shortstack{$\zeta^{-1}e^{-(\ln\zeta)^2/2s^2}$\\(log-normal)} & $0$ \\
\hline
\end{tabular}}
\end{table}

\subsubsection{Family A: symmetric generalized normal}
\label{subsec:famA}

The generalized normal (GN) distribution is the minimal one-parameter
deformation of the Gaussian and was introduced in this context by Ref.~\cite{SMBHs_direct_collapse_Nakama}; it was subsequently applied to the spectral-distortion problem by Hooper~\textit{et al.}~\cite{Hooper2024}, who used a symmetric distribution of this form to show that a sufficiently heavy tail evades the FIRAS bound at fixed variance. The distribution corresponds to a symmetric stretching redefinition of the Gaussian field, $\zeta = c\,\mathrm{sgn}(\delta\phi)\,|\delta\phi|^{2/p}$, which through Eq.~(\ref{eq:change_of_variables}) gives $P_\zeta \propto \exp(-|\zeta/c|^p)$. 

\begin{figure*}[t]
\centering
\subfigure[$M=10^6\,M_\odot$]{\includegraphics[width=3.1in]{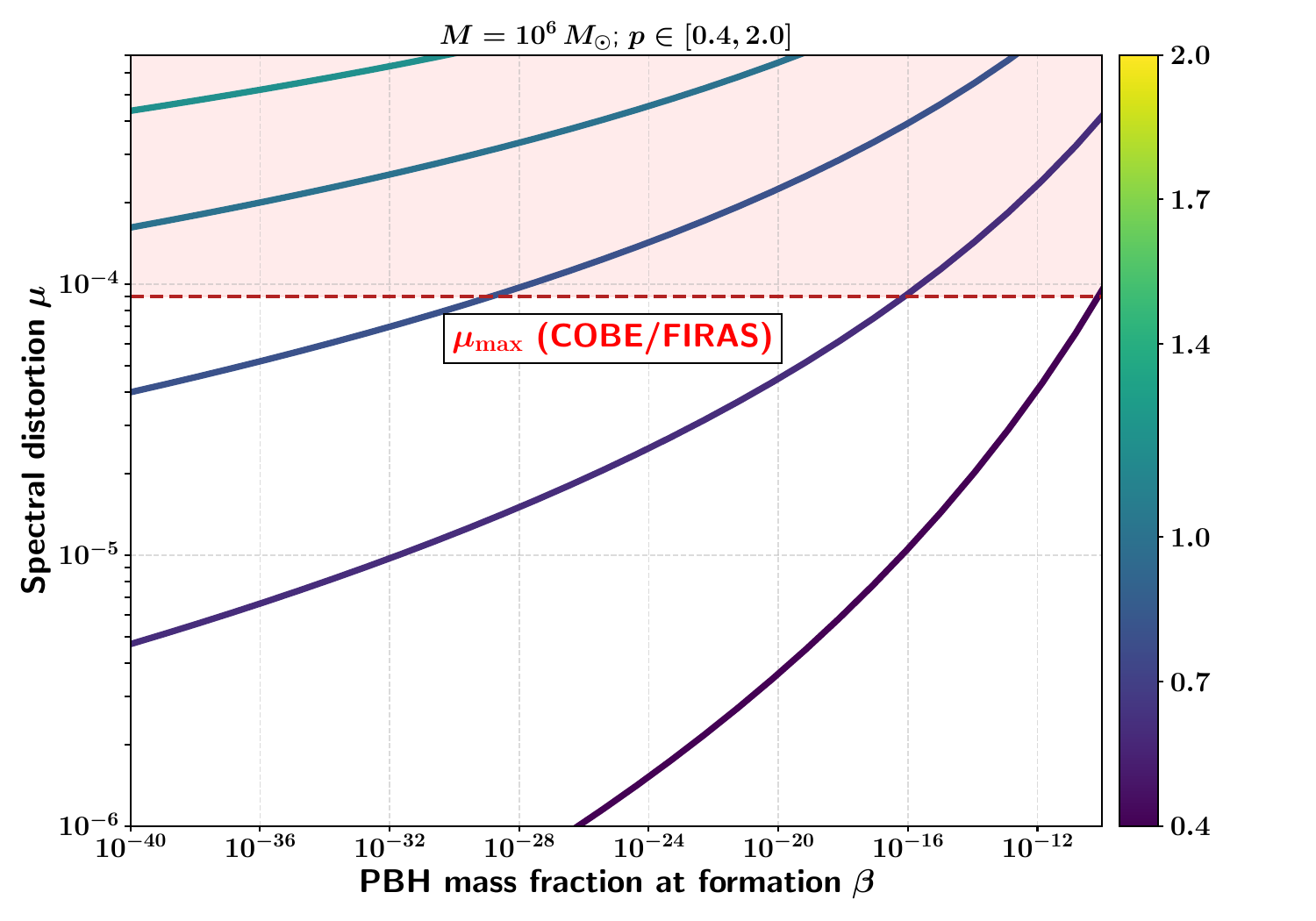}}
\hfill
\subfigure[$M=10^7\,M_\odot$]{\includegraphics[width=3.1in]{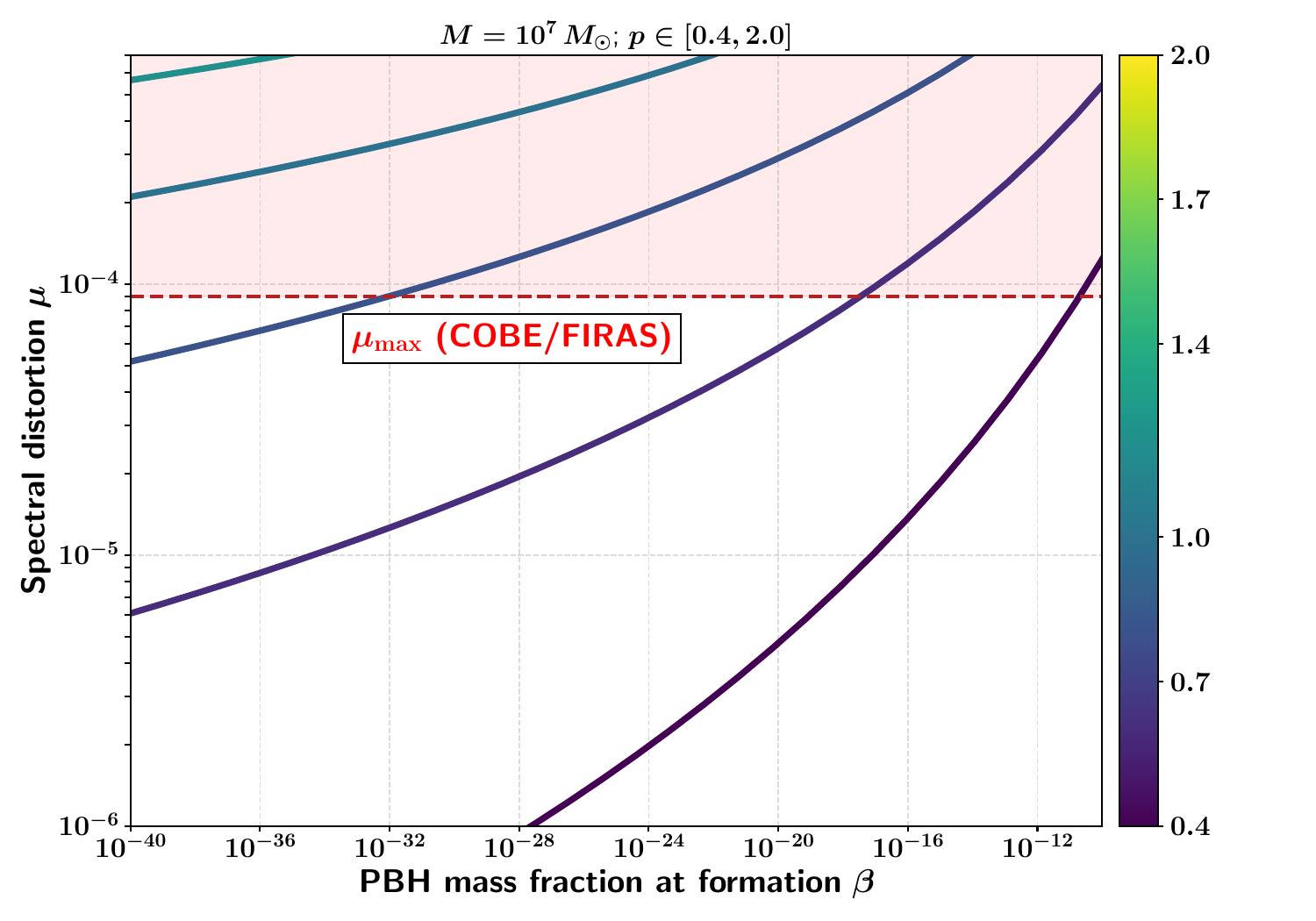}}
\caption{$\mu$-distortion versus PBH mass fraction $\beta$ at formation for the generalized-normal family (Family A), at $M=10^6\,M_\odot$ (left) and $10^7\,M_\odot$ (right). Each curve is a fixed $p$; moving along it varies the variance, and the vertical line marks the FIRAS limit $\mu<9\times10^{-5}$. As the seed mass grows the distortion-capped variance shrinks, so reaching a fixed $\beta$ within the bound requires progressively smaller $p$ (heavier tails).}
\label{fig:mu_vs_beta_GN}
\end{figure*}

 The standardized PDF is
\begin{equation}
\widehat{P}_\mathrm{GN}(u;p)=\frac{p}{2\,\lambda\,\Gamma(1/p)}\,
\exp\!\left(-\left|\frac{u}{\lambda}\right|^{p}\right),
\quad
\lambda\equiv\sqrt{\frac{\Gamma(1/p)}{\Gamma(3/p)}},
\label{eq:GN_standard}
\end{equation}
which recovers the Gaussian at $p=2$ and the Laplace (exponential) law at $p=1$; for $0<p<1$ the tail is heavier than exponential, $D_\infty=0$. Decreasing $p$ at fixed $\mu$-limited variance raises $\beta_{\max}$ by many orders of magnitude (Fig.~\ref{fig:mu_vs_beta_GN}): a smaller $p$ is a heavier symmetric tail and hence a larger collapse probability at fixed $u_c$. 

The GN family is, however, symmetric: it assigns equal heavy weight to deep voids ($\zeta<0$), which no single-field, one-sided non-attractor feature produces. We therefore use Family A only as a diagnostic that isolates how much tail weight is required, before turning to the asymmetric, physically motivated families B--D.

\subsubsection{Family B: Gaussian core with stretched-exponential tail}
\label{subsec:famB}

Single-field models with a transient non-attractor phase preserve Gaussianity in the perturbative core while developing a one-sided heavy tail. Their exact $\delta N$ map follows from the logarithmic duality of Pi \& Sasaki~\cite{Pi_2023}, who solve the constant-roll equation of motion on a piecewise-quadratic potential. With $\eta\equiv m^2/3H^2$, the characteristic exponents are
\begin{equation}
\lambda_\pm=\frac{3\pm\sqrt{9-12\eta}}{2}.
\label{eq:char_roots}
\end{equation}
When one logarithm dominates, the curvature perturbation reduces to a single logarithm of the Gaussian field,
\begin{equation}
\zeta=-\frac{1}{\lambda}\,\ln\!\left(1+c\,\delta\phi\right),
\label{eq:PiSasaki_map}
\end{equation}
with $c$ a model-dependent constant and $\lambda$ the dominant characteristic root. Because the logarithm saturates, the Gaussian argument approaches a finite value as $\zeta\to\infty$ while the Jacobian in Eq.~\eqref{eq:change_of_variables} dominates, producing an exponential tail $P_\zeta\propto e^{-\lambda\zeta}$; the value of $\lambda$ for the analytically solvable cases is tabulated in Ref.~\cite{Pi_2023}. The ultra-slow-roll case fixes $\lambda=3$ (the larger root for $\eta=0$) and $c=3/\pi$, i.e.\ $\zeta=-\tfrac13\ln(1+3\,\delta\phi/\pi)$, which gives the well-known exponential tail~\cite{AtalGermani2019,EzquiagaGarciaBellidoVennin2020}
\begin{equation}
P_\zeta(\zeta)\propto e^{-3\zeta},\qquad D_\infty=3>0.
\label{eq:exp_tail}
\end{equation}
The same conclusion holds for a finite-width upward step: accounting for the step width $\Delta\phi$, Ref.~\cite{Kawaguchi2023} finds $P_\zeta(\zeta)\propto e^{-2\omega_{s2}\zeta}$ with $\omega_{s2}\propto1/\Delta\phi$, so a narrower step gives a steeper (not heavier) tail and the zero-width hard cutoff is recovered as $\omega_{s2}\to\infty$. In summary, canonical single-field non-attractor transitions yield an exponential tail, $p=1$~\cite{Pi_2023,CaiEtAl2022,InuiEtAl2024}; by themselves they do not reach the sub-exponential regime $p<1$. We quantify this in Sec.~\ref{sec:Results}. (The lighter Gumbel/double-exponential case arises for non-attractor quadratic segments and is not modeled here.)

\begin{figure*}[t]
\centering
\subfigure[Standardized PDF]{\includegraphics[width=3.1in]{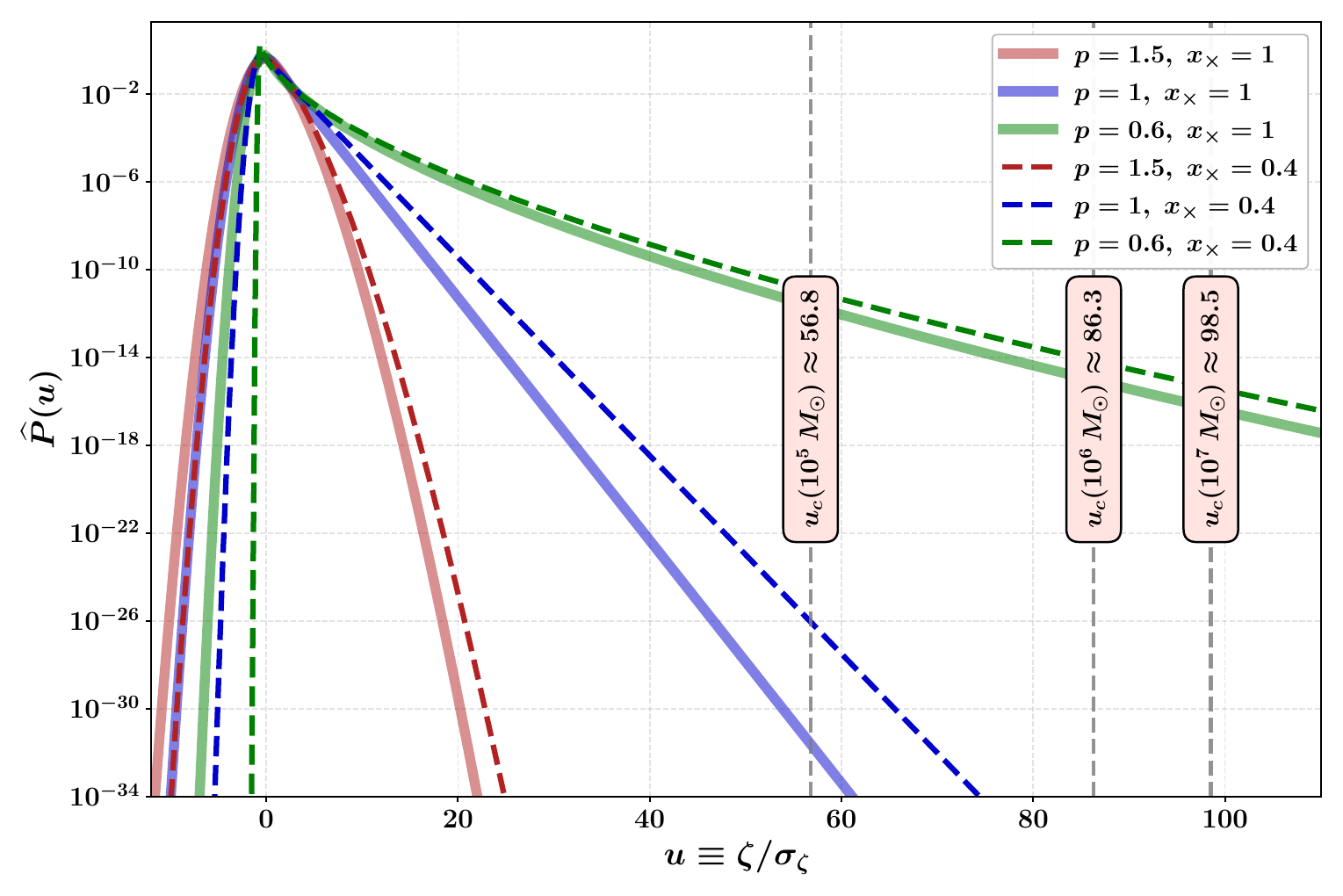}}
\hfill
\subfigure[$\beta_{\max}(M)$ at formation]{\includegraphics[width=3.1in]{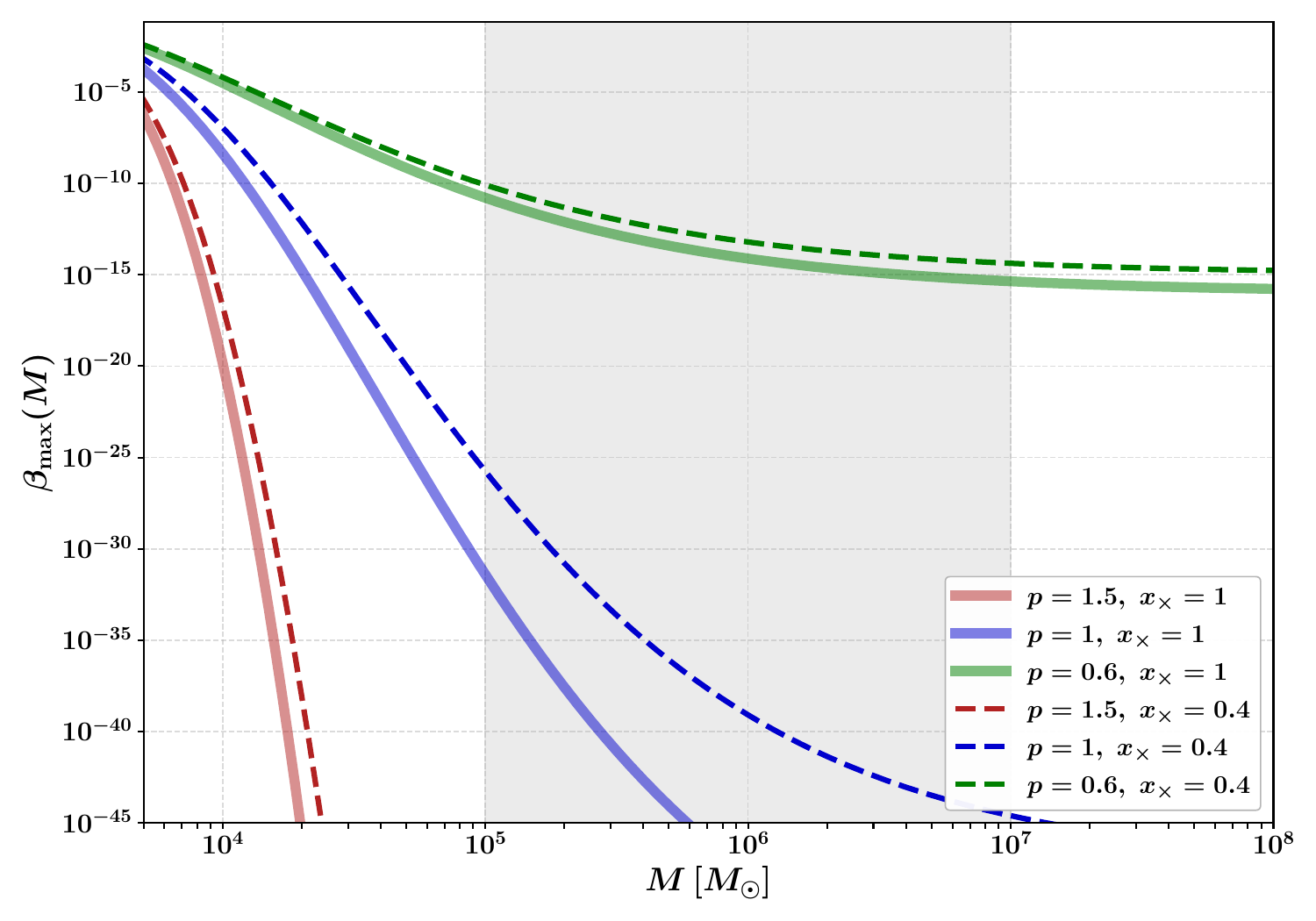}}
\caption{Family B (Gaussian core with stretched-exponential tail). \textbf{(a)} Standardized PDFs $\widehat P(u)$ and \textbf{(b)} distortion-capped $\beta_{\max}(M)$, for tail indices $p=1.5,1,0.6$ at two transition points, $x_\times=1$ (solid) and $x_\times=0.4$ (dashed). In panel (a) the labeled vertical lines mark the collapse thresholds $u_c(10^5)\simeq56.8$, $u_c(10^6)\simeq86.3$, and $u_c(10^7)\simeq98.5$; in panel (b) the shaded band is the seed window $10^{5}$--$10^{7}\,M_\odot$. The compressed-exponential case $p=1.5$ is lighter than exponential and falls off immediately, and the single-field value $p=1$ remains far below the seed-relevant range across the window, whereas the sub-exponential $p=0.6$ flattens near $\beta_{\max}\sim10^{-13}$--$10^{-15}$; a smaller $x_\times$ (earlier onset of the tail) raises $\beta_{\max}$ at fixed $p$. The viable $p$ is mapped in Fig.~\ref{fig:param_space_scans}.}
\label{fig:PDF_beta_max_stretched}
\end{figure*}

Family B parametrizes this class by matching a stretched-exponential tail of index $p$ to a Gaussian core at a transition $x_\times$,
\begin{equation}
P_\mathrm{ST}^\mathrm{raw}(x;p,x_\times)=\frac{1}{Z_\mathrm{ST}}
\begin{cases}
\exp\!\left(-\dfrac{x^2}{2}\right), & x\le x_\times,\\[10pt]
\exp\!\left[-\Phi_\mathrm{ST}(x)\right], & x>x_\times,
\end{cases}
\label{eq:stretched_raw}
\end{equation}
\begin{equation}
\Phi_\mathrm{ST}(x)=\frac{x_\times^{\,2-p}}{p}\,x^{p}
+x_\times^{2}\!\left(\frac{1}{2}-\frac{1}{p}\right),
\label{eq:phi_ST}
\end{equation}
where the form of $\Phi_\mathrm{ST}$ enforces continuity of the PDF and its first derivative at $x=x_\times$. The two parameters have the following interpretation: $p$ is the asymptotic tail index of Eq.~\eqref{eq:exp_tail} and its sub-exponential generalizations ($p=1$ being the value derived for the single-field USR/step case), while $x_\times$ is the transition point of the unstandardized PDF; its standardized counterpart $u_\times=(x_\times-\mu_x)/\sigma_x$ is a derived quantity and is not held fixed across the family. Figure~\ref{fig:PDF_beta_max_stretched} shows the resulting standardized PDFs and the distortion-capped abundance $\beta_{\max}(M)$.

\subsubsection{Family C: Gaussian core with power-law tail}
\label{subsec:famC}

 The heaviest physically motivated tails are algebraic and arise from
classical non-attractor dynamics rather than from single-clock stochastic diffusion, which on a flat plateau instead regulates fluctuations into exponential tails~\cite{Pattison2017,EzquiagaGarciaBellidoVennin2020}. We
work throughout at the level of the classical $\delta N$ map,
Eq.~(\ref{eq:deltaN_master}), with Gaussian field fluctuations, and read the resulting tail shapes as those of the underlying classical trajectory; the diffusive regime is outside our scope. Following
Refs.~\cite{Hooshangi2022,Hooshangi2023}, a power-law tail is obtained whenever the field displacement saturates to a finite value
$\delta\phi_\mathrm{max}$ at large $e$-fold number, so that the Gaussian fall-off terminates. Requiring the asymptotic behavior
\begin{equation}
\delta N\sim\big[\tilde\gamma\,(\delta\phi_\mathrm{max}-\delta\phi)\big]^{-1/q},
\label{eq:dN_powerlaw}
\end{equation}
with $\tilde\gamma$ a model-dependent constant (not to be confused with the collapse fraction $\gamma$ of Eq.~\eqref{Eq:mass_k_relation}), the Jacobian in Eq.~\eqref{eq:change_of_variables} yields $P_\zeta(\zeta)\propto\zeta^{-(q+1)}$ while $P_G\to\mathrm{const}$, i.e.\ $D_\infty=0$. A concrete potential realizes this behavior: integrating the background equation for the trajectory~\eqref{eq:dN_powerlaw} gives, in the de Sitter regime~\cite{Hooshangi2023},
\begin{equation}
V(\phi)=V_0\!\left[1+\frac{q^{2}}{2q+1}\,\tilde\gamma^{1/q}\,
\big|\phi-\bar\phi\big|^{\,2+1/q}\right],
\label{eq:fractional_V}
\end{equation}
so that the leading deviation from flatness is a fractional power $V-V_0\propto|\phi-\bar\phi|^{m}$ with
\begin{equation}
m=2+\frac{1}{q}
\qquad\Longleftrightarrow\qquad
q=\frac{1}{m-2}.
\label{eq:m_q_relation}
\end{equation}
A finite variance, which is a prerequisite for the spectral-distortion bound to apply, requires $q>2$, i.e.\ $2<m<2.5$; the broader range $2<m<3$ gives $q>1$ but includes variance-divergent tails, which we exclude. Fractional powers of this kind are the natural effective description of radiatively corrected inflection points, $V\sim\phi^{4}\ln(\phi/\mu)$~\cite{Barenboim_2017}, and arise when a smooth potential is expressed in the canonical field frame of a non-canonical (e.g.\ Dirac--Born--Infeld) kinetic sector~\cite{SilversteinTong2004,AlishahihaSilversteinTong2004,Dirac_Born_Infled_SSR}; we adopt Eq.~\eqref{eq:fractional_V} as the representative classical realization and leave an explicit Dirac--Born--Infeld construction to future work.

\begin{figure*}[t]
\centering
\subfigure[Standardized PDF]{\includegraphics[width=3.1in]{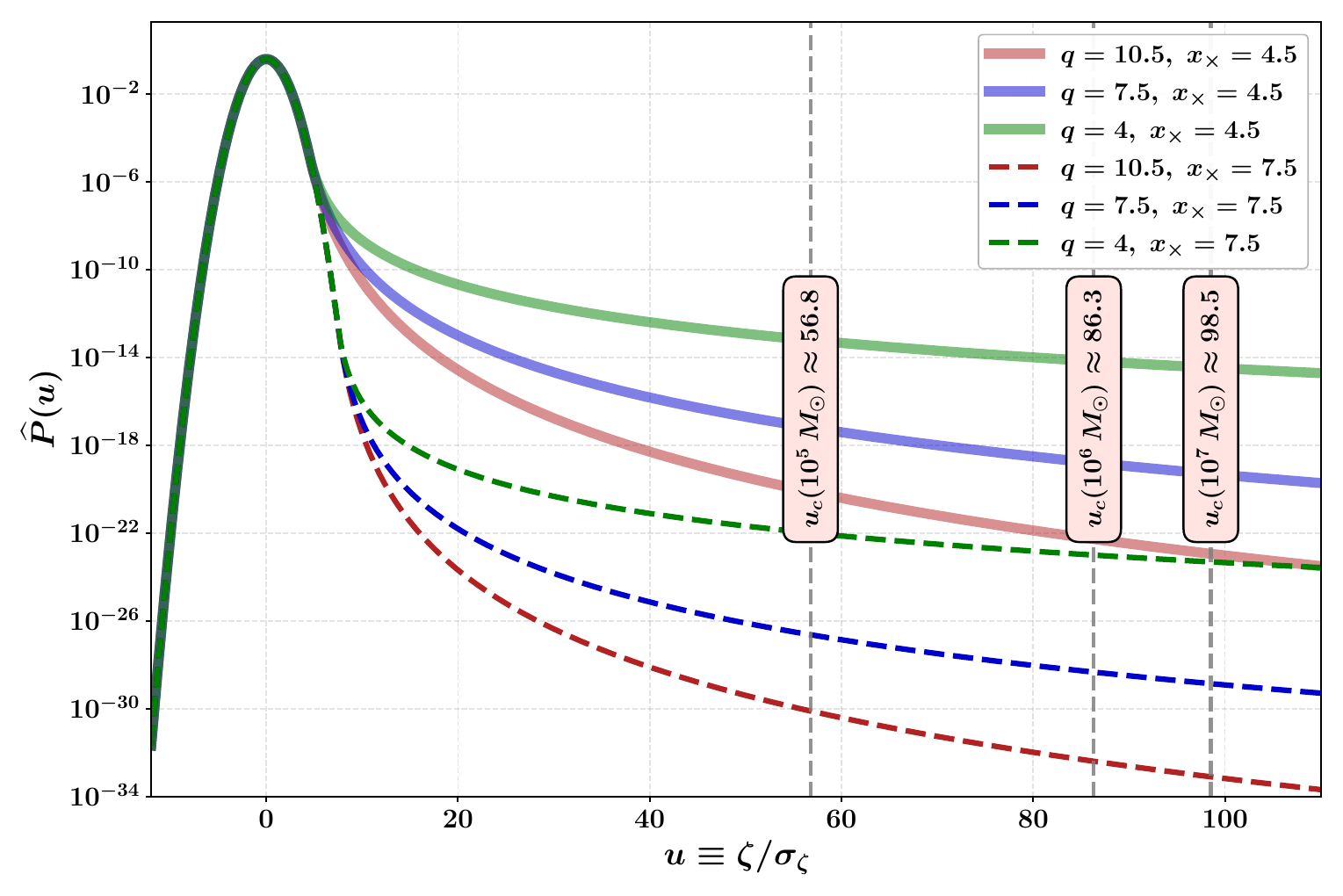}}
\hfill
\subfigure[$\beta_{\max}(M)$ at formation]{\includegraphics[width=3.1in]{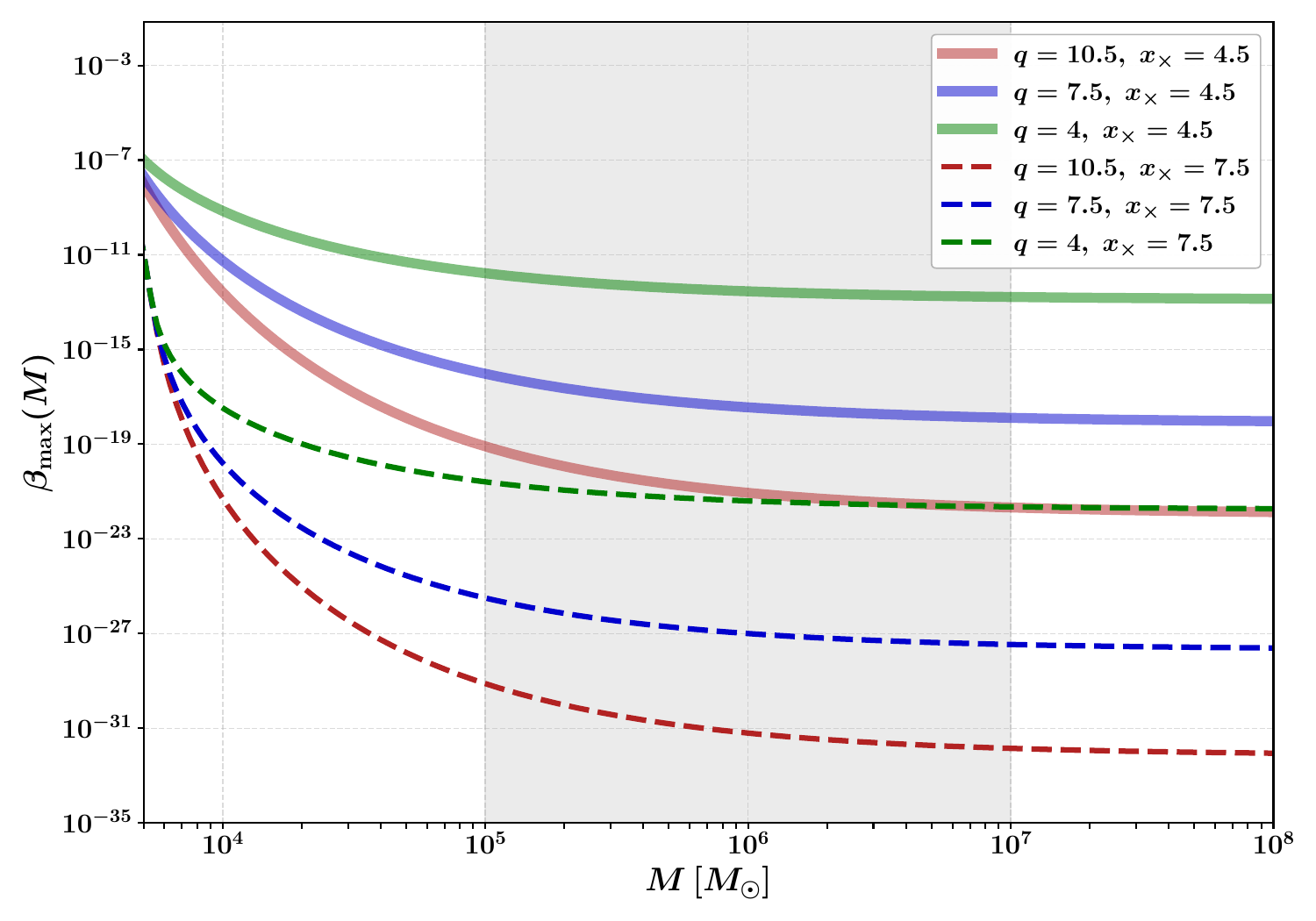}}
\caption{Family C (Gaussian core with power-law tail). \textbf{(a)} Standardized PDFs $\widehat P(u)$ and \textbf{(b)} distortion-capped $\beta_{\max}(M)$, for indices $q=10.5,7.5,4$ (with $q>2$ for finite variance) at two transition points, $x_\times=4.5$ (solid) and $x_\times=7.5$ (dashed). Vertical lines and the shaded band are as in Fig.~\ref{fig:PDF_beta_max_stretched}. Because the algebraic tail carries no exponential suppression, $\beta_{\max}(M)$ flattens toward a nearly mass-independent floor that rises as $q$ and $x_\times$ decrease; the family is far less sensitive to the seed mass than the stretched family.}
\label{fig:PDF_beta_max_PL}
\end{figure*}

Family C matches this algebraic tail to a Gaussian core,
\begin{equation}
P_\mathrm{PL}^\mathrm{raw}(x;q,x_\times)=\frac{1}{Z_\mathrm{PL}}\times
\begin{cases}
e^{-x^2/2}, & x\le x_\times,\\[6pt]
\begin{aligned}
&e^{-x_\times^2/2}\\
&\times\left(\dfrac{x-x_0}{x_\times-x_0}\right)^{-(q+1)},
\end{aligned}
& x>x_\times.
\end{cases}
\label{Eq:PowerLawPDF}
\end{equation}
with $x_0\equiv x_\times-(q+1)/x_\times$ enforcing $C^1$ continuity at $x_\times$ and $q>2$ imposed for finite variance. As anticipated by the $D_\infty=0$ classification, the algebraic decay carries no exponential suppression at the large thresholds $u_c\sim50$--$100$, and therefore dominates any stretched tail with $p\gtrsim0.6$ (Fig.~\ref{fig:PDF_beta_max_PL}). The index $q$ is fixed by the potential via Eq.~\eqref{eq:m_q_relation}, and $x_\times$ marks the onset of the non-perturbative branch.

\subsubsection{Family D: asymmetric log-normal tail}
\label{subsec:famD}

A log-normal tail arises from the \emph{inverse} structure, in which the physical variable is the exponential of a near-Gaussian field, $1+\kappa\zeta=e^{sX}$. This is the multiplicative analogue of the central limit theorem. A quantity assembled as a product of many independent factors, $\prod_i(1+\epsilon_i)$, has a logarithm that is a sum of independent terms, hence asymptotically Gaussian; the quantity itself is therefore log-normal~\cite{Limpert2001}. Multiplicative structure of this kind is familiar in cosmology. Coles \& Jones~\cite{ColesJones1991} model the nonlinear matter density as log-normal, $1+\delta\simeq\exp(\delta_{\rm lin}-\sigma^{2}/2)$, precisely because density accumulates multiplicatively. The same exponential structure enters PBH formation through the nonlinear relation between the collapse variable (the density contrast, or compaction) and $\zeta$ on super-horizon scales, which carries factors of $e^{\zeta}$ from the local volume element~\cite{YoungMuscoByrnes2019,YangCompaction2024}; a Gaussian $\zeta$ thus already maps to a multiplicatively skewed collapse variable.

\begin{figure*}[t]
\centering
\subfigure[Standardized PDF]{\includegraphics[width=3.1in]{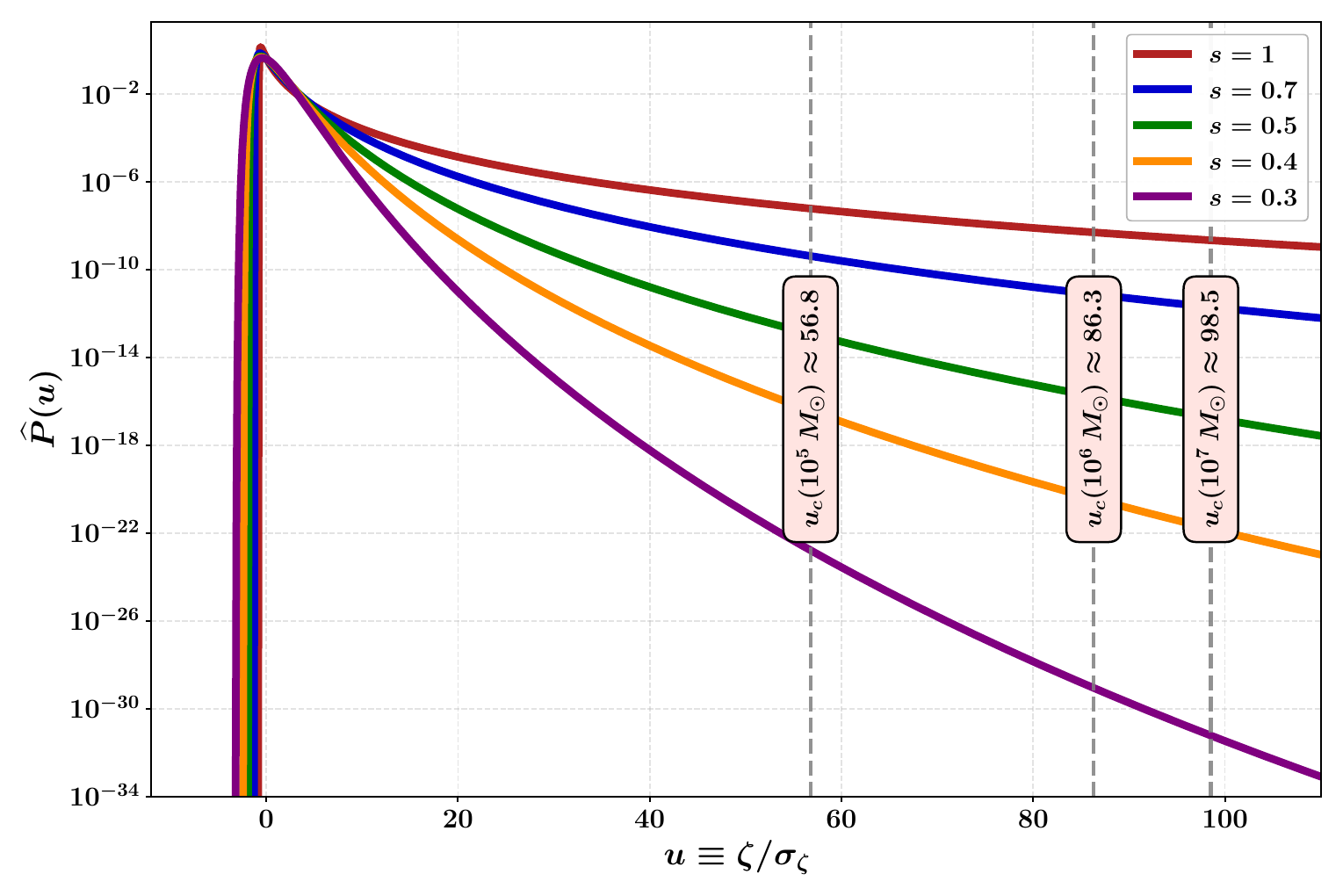}} 
\hfill
\subfigure[$\beta_{\max}(M)$ at formation]{\includegraphics[width=3.1in]{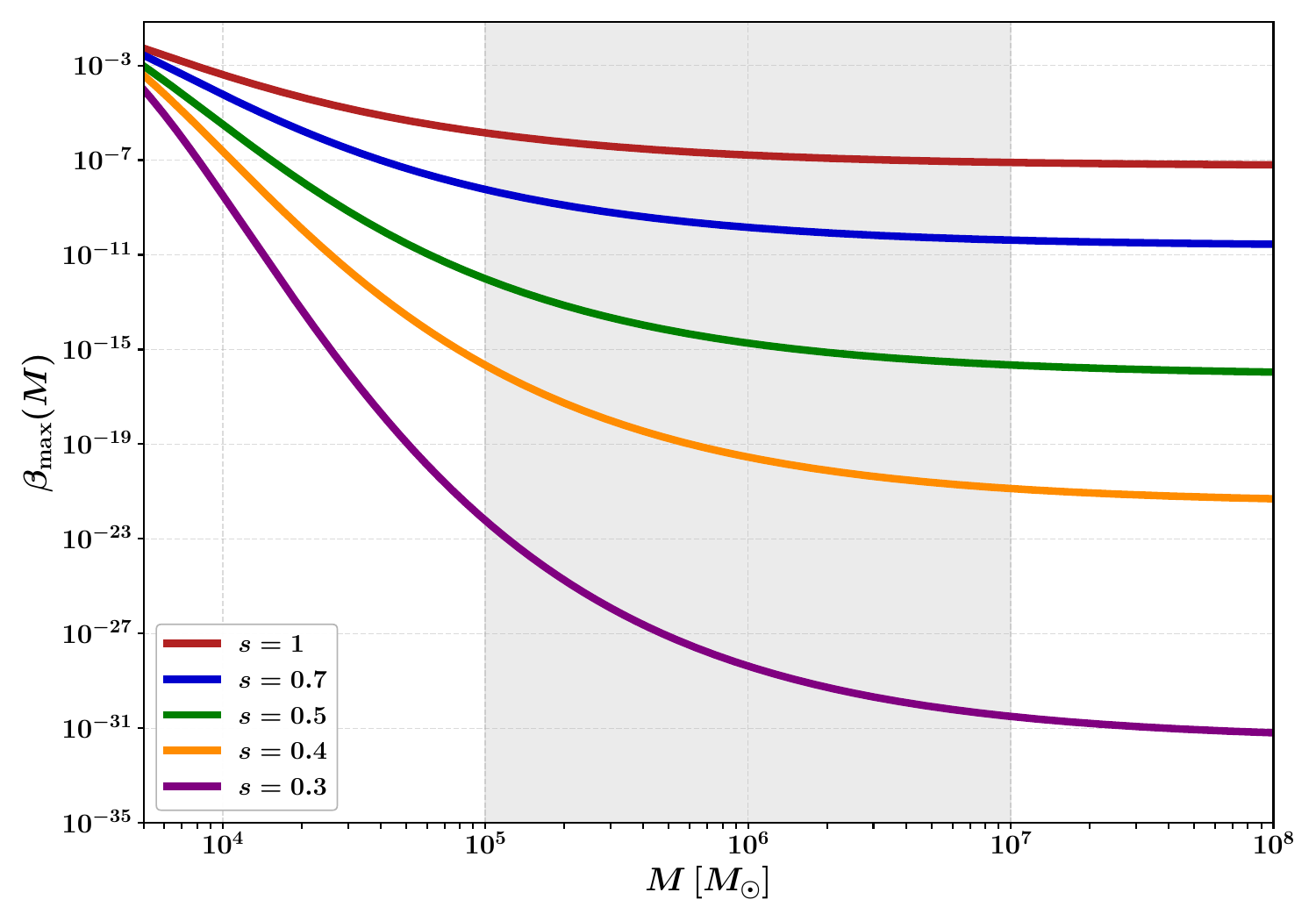}}
\caption{Family D (asymmetric log-normal). \textbf{(a)} Standardized PDF $\widehat P(u)$ and \textbf{(b)} distortion-capped $\beta_{\max}(M)$, for shape parameters $s=1,0.7,0.5,0.4,0.3$. Vertical lines and the shaded band are as in Fig.~\ref{fig:PDF_beta_max_stretched}. The log-normal is the heaviest family considered: for $s\gtrsim0.5$ its floor reaches the reference level $\beta\sim10^{-15}$ or above across the window, while the strong dependence on $s$ (spanning $\sim10^{-7}$ at $s=1$ down to $\sim10^{-31}$ at $s=0.3$) reflects the provisional, multiplicative origin discussed in Sec.~\ref{subsec:famD}.}
\label{fig:PDF_beta_max_lognormal}
\end{figure*}

In the inflationary scenario, however, single-clock stochastic dynamics does not reach this regime: the first-passage analysis of quantum diffusion yields exponential tails, consistent with the classification of Sec.~\ref{subsec:famC}~\cite{Pattison2017,EzquiagaGarciaBellidoVennin2020}. A log-normal tail instead requires the amplitude of the curvature perturbation to be built up multiplicatively, as can occur in multifield dynamics in which a second light field modulates the local expansion history~\cite{PanagopoulosSilverstein2020,VenninStarobinsky2015,PattisonVenninWandsAssadullahi2019}, or through the volume-weighted expansion factor $e^{3\delta N}$. We note that the map~\eqref{eq:LN_map} below is the functional inverse of the logarithmic map~\eqref{eq:PiSasaki_map}: where a logarithmic $\delta N$ map produces an exponential tail, an exponential $\delta N$ map produces a log-normal one, and the open question is which dynamics realizes the exponential map. We therefore treat Family D as the phenomenological representative of this multiplicative regime, the heaviest of our families, rather than as the output of a specific derived $\delta N$ map, and we return to its more provisional status in Sec.~\ref{sec:conclusion}.

Concretely, we take the shifted variable to be exactly log-normal,
\begin{equation}
\begin{aligned}
1+\kappa\zeta&=\exp\!\left(s\,X-\tfrac{s^2}{2}\right),\qquad X\sim\mathcal{N}(0,1),\\
&\Longleftrightarrow\quad
\ln(1+\kappa\zeta)\sim\mathcal{N}\!\left(-\tfrac{s^2}{2},s^2\right),
\end{aligned}
\label{eq:LN_map}
\end{equation}
where the centering $-s^2/2$ enforces $\langle\zeta\rangle=0$. The Jacobian transformation gives the raw PDF
\begin{equation}
\begin{aligned}
P_\mathrm{LN}^\mathrm{raw}(x;s,\kappa)=\;&
\frac{\kappa}{\sqrt{2\pi}\,s\,(1+\kappa x)}\\
&\times\exp\!\left[-\frac{\big(\ln(1+\kappa x)+\tfrac{s^2}{2}\big)^2}{2s^2}\right],
\end{aligned}
\label{eq:LN_raw}
\end{equation}
valid for $x>-1/\kappa$, whose positive tail, $P_\mathrm{LN}\sim\zeta^{-1}\exp[-(\ln\zeta)^2/2s^2]$, decays more slowly than any $e^{-\zeta^{n}}$ with $n>0$ (again $D_\infty=0$). The distribution is bounded below at $\zeta=-1/\kappa$, so it produces no unphysical deep voids, and after standardization to unit variance it reduces to a one-parameter family in $s$ (Fig.~\ref{fig:PDF_beta_max_lognormal}).

\subsection{Standardization and the collapse integral}
\label{subsec:pipeline}

We summarize the procedure that maps any raw family above to a distortion-capped abundance, taking Family C as the worked example. The mass fraction at formation is the Press--Schechter integral~\cite{Press_Schechter},
\begin{equation}
\beta(M)=2\int_{\zeta_c}^{\infty}P_\zeta(\zeta\,|\,M)\,\mathrm{d}\zeta,
\qquad \zeta_c\simeq0.67.
\label{eq:beta_PS}
\end{equation}
First, the FIRAS bound caps the variance, Eq.~\eqref{eq:sigma_cap_mu}, which we saturate; at $M=10^{6}\,M_\odot$, $W_\mu\approx0.679$ and $\sigma_{\zeta,\max}^{2}\approx6.0\times10^{-5}$ ($\sigma_{\zeta,\max}\approx7.8\times10^{-3}$). Second, because attaching a heavy asymmetric tail shifts the raw mean and variance, we standardize numerically: with
\begin{align}
\mu_x&=\frac{1}{Z}\int_{-\infty}^{\infty}x\,P^\mathrm{raw}(x)\,\mathrm{d}x,
\label{eq:raw_mean}\\
\sigma_x^2&=\frac{1}{Z}\int_{-\infty}^{\infty}x^2\,P^\mathrm{raw}(x)\,\mathrm{d}x-\mu_x^2,
\label{eq:raw_var}
\end{align}
we set $u=(x-\mu_x)/\sigma_x$, so that $\widehat{P}(u)=\sigma_x\,P^\mathrm{raw}(\mu_x+\sigma_x u)$ has zero mean and unit variance. The physical and standardized variables are related by $\zeta=\sigma_{\zeta,\max}(M)\,u$, which locks the core variance to the distortion bound while leaving the tail shape free.

The abundance then follows from Eq.~\eqref{eq:beta_max} with the collapse function evaluated at $u_c(M)$. For the $10^{6}\,M_\odot$ example $u_c\approx86$, where a Gaussian gives $\mathcal{S}\sim e^{-u_c^{2}/2}\sim10^{-1600}$, while the standardized power-law tail gives $\mathcal{S}_\mathrm{PL}\sim u_c^{-q}\sim10^{-7}$ for $q=3.5$. The algebraic decay thus raises the collapse probability by more than $1500$ orders of magnitude at fixed variance.

\section{Results}
\label{sec:Results}

We now evaluate the distortion-capped abundance $\beta_{\max}(M)$ of Eq.~\eqref{eq:beta_max} for the four tail families. The FIRAS bound is saturated at every mass, so the standardized collapse threshold grows from $u_c\simeq57$ at $10^{5}\,M_\odot$ to $u_c\simeq99$ at $10^{7}\,M_\odot$, and no parameter is tuned to a preferred abundance. The curves show the full scanned range with $\beta \sim 10^{-15}$ marked as a useful reference corresponding to $f_{\rm PBH}\sim10^{-10}$ across the seed window. 

Three approximations bound the scope of what follows, and none of them affects the ordering of the families. The narrow-feature formula Eq.~\eqref{eq:mu_narrow} sets the variance cap; a broad spectral enhancement would instead require the full integral Eq.~\eqref{Eq:mu_window} evaluated on the model spectrum, shifting the cap by an $\mathcal{O}(1)$ factor at fixed mass. The Press--Schechter criterion with a fixed threshold on $\zeta$ simplifies the collapse condition; we test the threshold dependence below, and a treatment based on the compaction function and critical collapse would refine the absolute abundances without reversing a comparison in which exponential and sub-exponential tails differ by tens of orders of magnitude. Finally, the log-normal family is phenomenological at the level of the $\delta N$ map, and we read it throughout as indicative of the heaviest, multiplicative regime rather than as a derived prediction.

\begin{figure*}[htbp]
\centering
\subfigure[]{\includegraphics[width=3.3in]{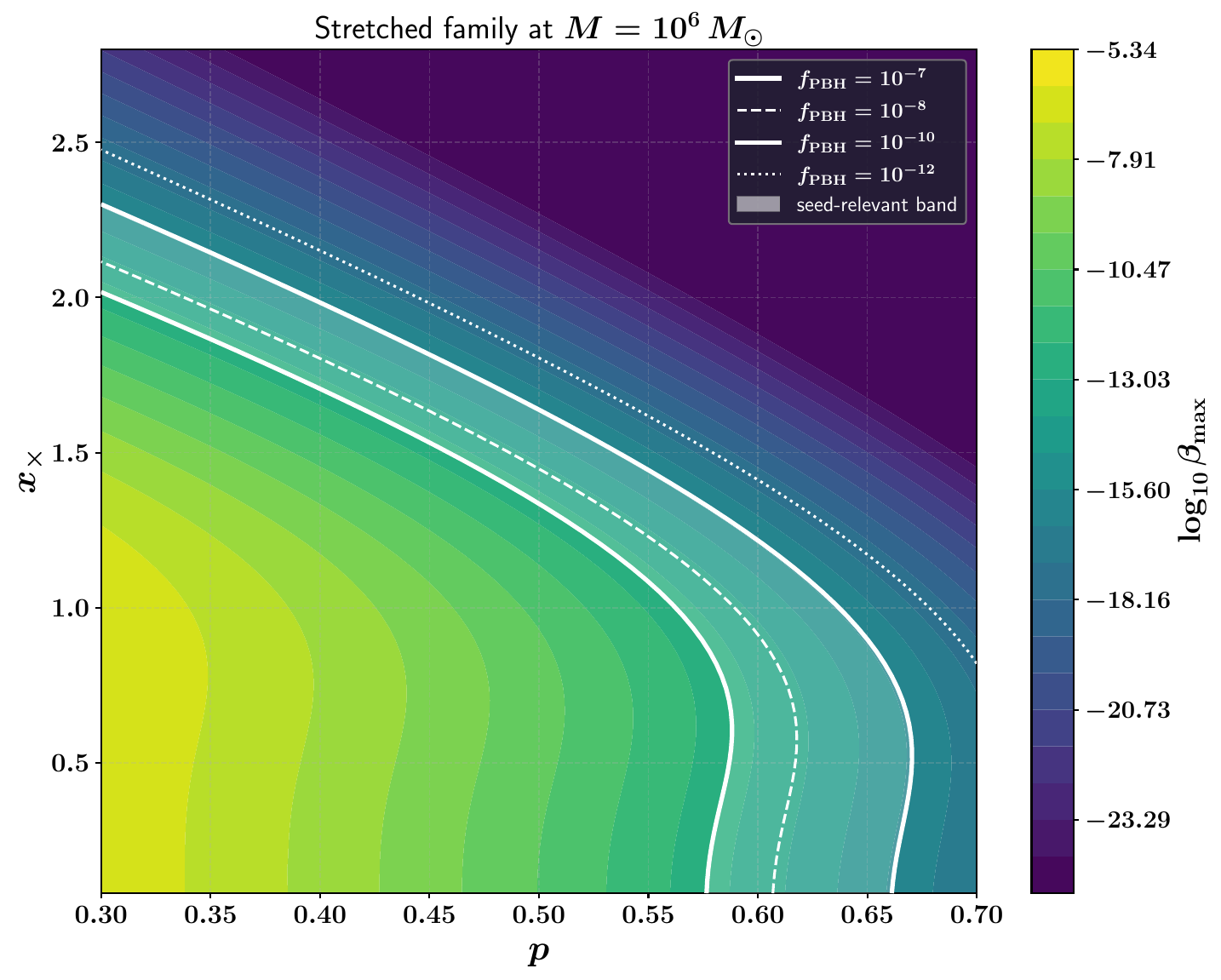}}
\hfill
\subfigure[]{\includegraphics[width=3.3in]{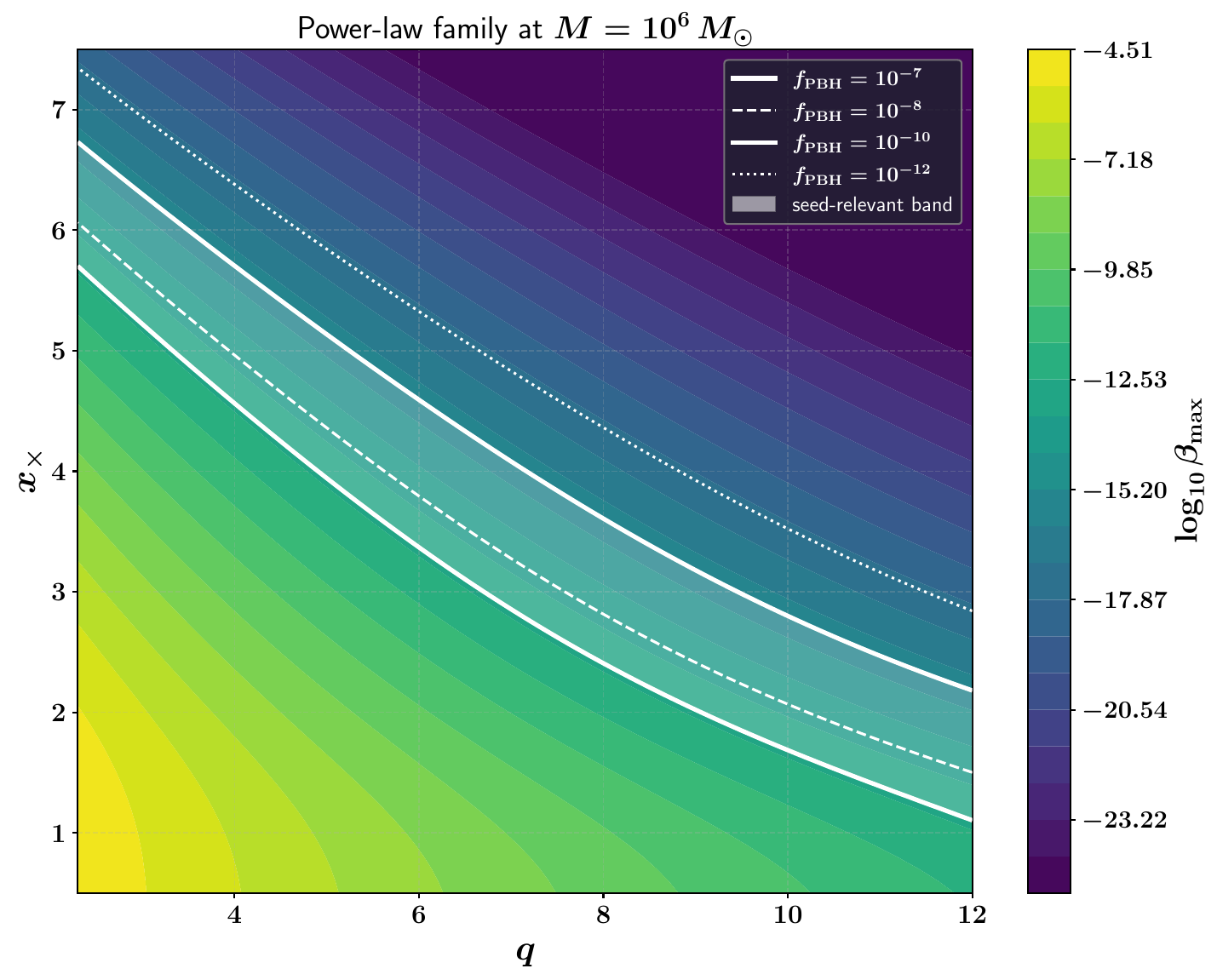}}\\
\subfigure[]{\includegraphics[width=3.3in]{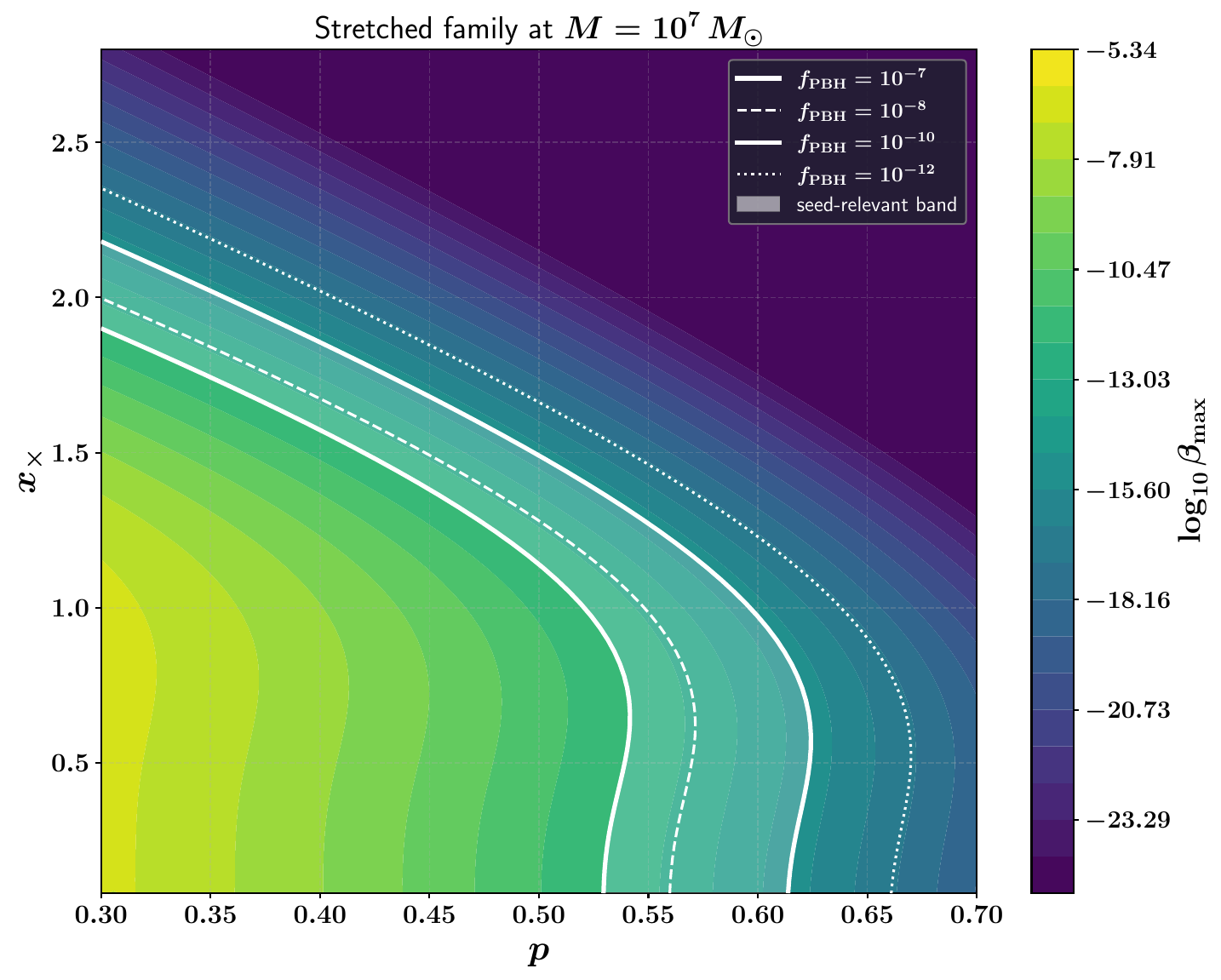}}
\hfill
\subfigure[]{\includegraphics[width=3.3in]{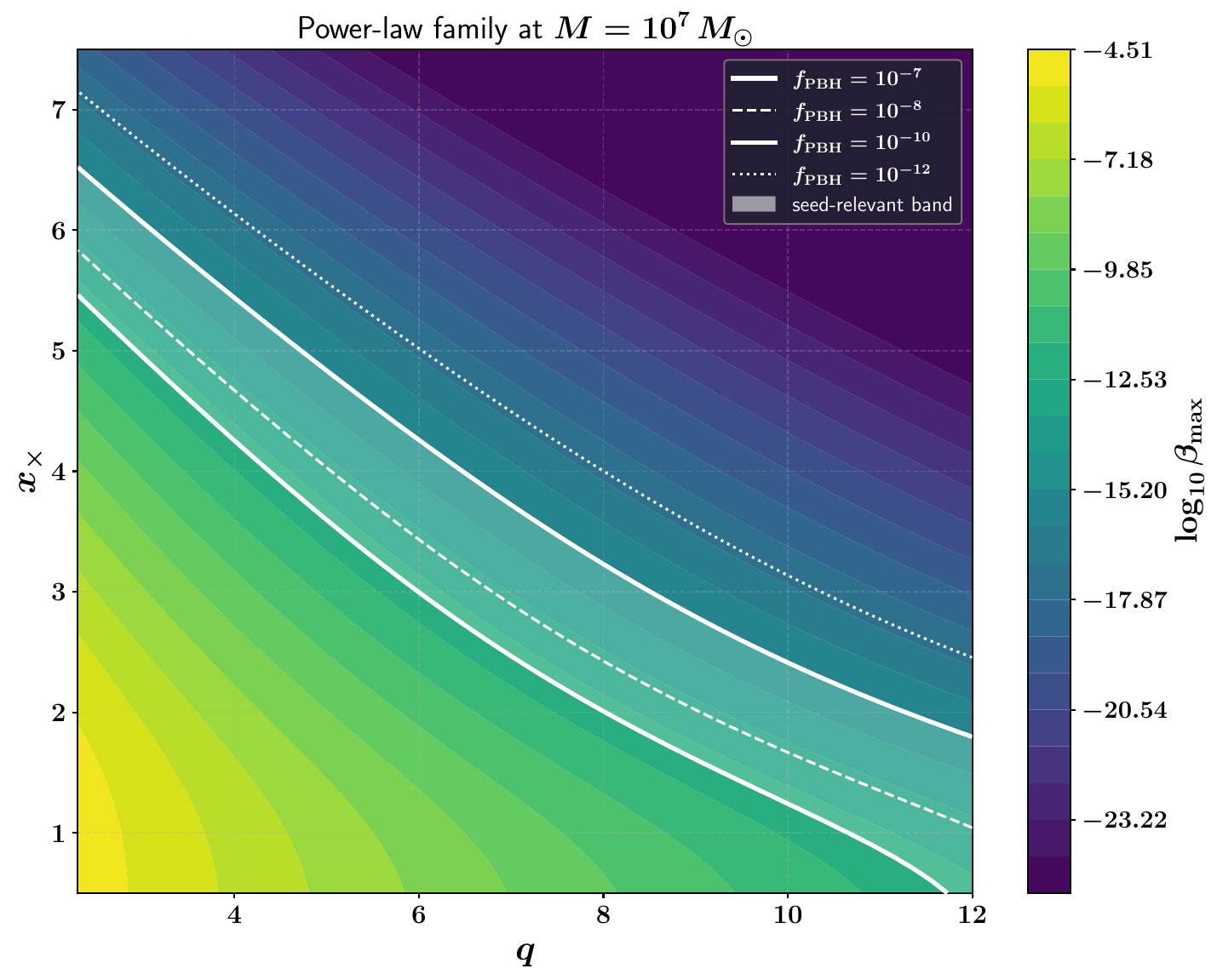}}
\caption{Distortion-capped abundance for the stretched-exponential (left) Eq.~\eqref{eq:stretched_raw} and power-law (right) Eq.~\eqref{Eq:PowerLawPDF} families in the $(p,x_\times)$ and $(q,x_\times)$ planes, at $M=10^{6}\,M_\odot$ (top) and $10^{7}\,M_\odot$ (bottom). The color map shows $\log_{10}\beta_{\max}$ at the saturated $\mu$-distortion cap; white lines are iso-contours of the present-day fraction $f_{\rm PBH}=10^{-7},10^{-8},10^{-10},10^{-12}$, obtained from $\beta_{\max}$ through the mass-dependent relation Eq.~\eqref{eq:beta_to_f}; the shaded band between $f_{\rm PBH}=10^{-10}$ and $10^{-7}$ marks the range relevant for seeding (Sec.~\ref{subsec:results_limits}). The scan variable $x_\times$ is the transition point of the unstandardized PDF (Sec.~\ref{sec:PDF_analyses}). The minimal one-seed requirement lies well below the band, while its upper portion ($f_{\rm PBH} \gtrsim 10^{-9}$) is in tension with the most aggressive accretion limits; both are discussed in Sec.~\ref{subsec:results_limits}. }
\label{fig:param_space_scans}
\end{figure*}

\subsection{Tail families and inflationary dynamics}
\label{subsec:results_phys}

Figures~\ref{fig:PDF_beta_max_stretched}--\ref{fig:PDF_beta_max_lognormal} present $\beta_{\max}(M)$ for the stretched-exponential, power-law, and log-normal families. 
Once the FIRAS bound fixes the variance, the abundance is controlled almost entirely by the far tail at $u_c\sim50$--$100$. The Gaussian core determines the typical fluctuations but is irrelevant for collapse. A tail with $D_\infty>0$ keeps an exponential penalty at the threshold; a sub-exponential tail with $D_\infty=0$ removes that penalty and can survive the tightening of the variance cap toward larger seed masses.

A Gaussian gives $\beta_{\max}\sim10^{-1600}$ at $10^{6}\,M_\odot$, while the ordinary exponential tail, $p=1$, raises the abundance by more than a thousand orders of magnitude. It remains below $\beta_{\max}\sim10^{-30}$ at the same mass and declines rapidly across the seed window. The viable stretched-exponential cases lie instead near $p\simeq0.6$, where $\beta_{\max}$ stays close to the reference level $\beta\sim10^{-15}$ across the band. This threshold is consistent with fixed-order non-Gaussianity studies: the quadratic ($f_{\rm NL}$, $p=1$) and cubic ($g_{\rm NL}$, $p=2/3$) maps remain too light~\cite{UnalKovetzPatil2021,Byrnes:2024vjt,Pritchard:2025yda}, while $\zeta\propto\zeta_g^{5}$ ($p=0.4$) can generate sub-exponential tails, which can reopen the seed window~\cite{Byrnes:2024vjt,Pritchard:2025yda}. 

To locate the viable region we scan the shape parameters at fixed mass. Figure~\ref{fig:param_space_scans} maps $\log_{10}\beta_{\max}$ over the $(p,x_\times)$ plane of the stretched-exponential family Eq.~\eqref{eq:stretched_raw} and the $(q,x_\times)$ plane of the power-law family Eq.~\eqref{Eq:PowerLawPDF}, at $M=10^{6}$ and $10^{7}\,M_\odot$, with iso-contours of the present-day fraction $f_{\rm PBH}$ from Eq.~\eqref{eq:beta_to_f}. The onset $x_\times$ is the transition point of the unstandardized PDF and enters the standardized threshold through Sec.~\ref{subsec:pipeline}. The $\mu$-distortion bound does not appear in these planes, because it has already fixed the variance; every contour is then placed by the tail shape alone; this makes the decoupling of Sec.~\ref{sec:background} explicit.

For the power-law family, $\mathcal{S}_\mathrm{PL}(u_c)\sim u_c^{-q}$, so the abundance is nearly insensitive to the modest change of $u_c$ across $10^{5}$--$10^{7}M_\odot$. The floor is set mainly by $q$ and by the onset $x_\times$; the heaviest case shown, $q=4$ and $x_\times=4.5$, remains above the seed reference throughout the window. This behavior is compatible with the fractional-potential branch of Eq.~\eqref{eq:fractional_V}, but it requires a genuine non-perturbative trajectory rather than a small deformation of slow roll. The log-normal family is even heavier, clearing the reference level for $s\gtrsim0.5$. Given its strong sensitivity to $s$ and the absence of a derived single-field $\delta N$ map, we treat it as a phenomenological parameterization of multiplicative or multifield dynamics~\cite{PanagopoulosSilverstein2020,VenninStarobinsky2015,Pattison2017,PattisonVenninWandsAssadullahi2019,Lin:2021vwc}, not as a prediction.

Each family represents a class of $\delta N$ maps (Table~\ref{tab:dictionary}), so these abundances carry implications for the underlying inflationary dynamics. The exponential tail $p=1$ is the generic output of single-field non-attractor transitions~\cite{CaiEtAl2022,CaiOneSmallStep2022,Kawaguchi2023,Pi_2023}, and Fig.~\ref{fig:PDF_beta_max_stretched} places it far below the seed range throughout. Such transitions can seed lighter PBHs~\cite{germani2017primordialblackholesinflection,Mishra_2020,Hertzberg_2018,Cicoli_2018}, but they do not reach the sub-exponential regime that SMBH seeds require.

The two families that do reopen the window point to more specific dynamics. The algebraic tail of Family~C follows from a fractional inflaton potential, Eq.~\eqref{eq:fractional_V} with $2<m<2.5$ for finite variance, of the form that arises as the effective description of radiatively corrected inflection points and of non-canonical (DBI) kinetic sectors~\cite{Barenboim_2017,SilversteinTong2004,AlishahihaSilversteinTong2004,Dirac_Born_Infled_SSR}. The log-normal tail of Family~D instead requires the curvature perturbation to accumulate multiplicatively~\cite{PanagopoulosSilverstein2020,VenninStarobinsky2015,Pattison2017,PattisonVenninWandsAssadullahi2019}, and remains the most provisional case. We do not claim that any published model attains these tails at the required amplitude. Our point is that the seed window selects a narrow class of dynamics, fractional-potential or multiplicative, and excludes the generic single-field exponential.

\subsection{Required abundance and astrophysical limits}
\label{subsec:results_limits}

The abundance required to seed the observed black holes is small, and it is set by simple number-density bookkeeping: a population of seeds of mass $M$ and comoving number density $n$ contributes $f_{\rm PBH}=n\,M/\rho_{\rm DM}$, with $\rho_{\rm DM}\simeq3.3\times10^{10}\,M_\odot\,{\rm Mpc}^{-3}$. Matching one seed to each luminous $z>6$ quasar, $n\sim10^{-9}\,{\rm Mpc}^{-3}$, gives $f_{\rm PBH}\sim3\times10^{-14}$ at $M=10^{6}\,M_\odot$ ($\beta\sim10^{-19}$). Demanding instead that every SMBH descend from a primordial seed raises $n$ to the local SMBH density and the requirement to $f_{\rm PBH}\sim10^{-7}$. The seed-relevant range therefore spans $f_{\rm PBH}\sim10^{-14}$--$10^{-7}$; the band shaded in Fig.~\ref{fig:param_space_scans}, $10^{-10}$--$10^{-7}$, sits within it, conservatively above the bare-quasar floor. Table~\ref{tab:seeding} lists the one-seed-per-quasar requirement across the seed-mass range.

\begin{table}[t]
\centering
\renewcommand{\arraystretch}{1.3}
\caption{One-seed-per-quasar requirement across the seed-mass range, from $f_{\rm PBH}=n\,M/\rho_{\rm DM}$ with $n\sim10^{-9}\,{\rm Mpc}^{-3}$ and $\rho_{\rm DM}\simeq3.3\times10^{10}\,M_\odot\,{\rm Mpc}^{-3}$, and the corresponding formation fraction $\beta$ from Eq.~\eqref{eq:beta_to_f}. The requirement rises with mass at fixed $n$ because each seed carries more mass. The whole-population case ($f_{\rm PBH}\sim10^{-7}$) uses the mass-dependent local SMBH density and is discussed in the text. }
\label{tab:seeding}
\begin{tabular}{@{}ccc@{}}
\hline\hline
$M\,[M_\odot]$ & $f_{\rm PBH}^{\rm req}$ (one per quasar) & $\beta^{\rm req}$ \\
\hline
$10^{5}$ & $3.0\times10^{-15}$ & $4.3\times10^{-21}$ \\
$10^{6}$ & $3.0\times10^{-14}$ & $1.4\times10^{-19}$ \\
$10^{7}$ & $3.0\times10^{-13}$ & $4.3\times10^{-18}$ \\
\hline\hline
\end{tabular}
\end{table}

Upper limits on $f_{\rm PBH}$ at these masses are heterogeneous and model dependent rather than a single sharp bound. Accretion limits, derived from the imprint of early accretion on CMB recombination and from present-day X-ray and radio emission, span several orders of magnitude depending on the assumed accretion geometry, from conservative spherical estimates~\cite{AliHaimoudKamionkowski2017} to more stringent disk analyses~\cite{Poulin2017,SerpicoPoulin2020,InoueKusenko2017}, and may relax further once ionization-front and feedback effects are included. Dynamical-friction and tidal arguments~\cite{CarrKuhnelVisinelli2021} and the Poisson (large-scale-structure) bound~\cite{CarrSilk2018} add independent and comparably uncertain constraints. The small-scale power spectrum itself is further constrained by the formation of ultracompact minihalos, via CMB bounds on their accretion~\cite{Croon:2024rmw,Bringmann:2025cht}; these can be stronger than the distortion bound over part of the seed window, and, being variance-type constraints, they would tighten $\sigma_{\zeta,\max}$ without affecting the comparison of tail shapes. Taken together with the compilations of Refs.~\cite{Carr2026,CarrKuhnelVisinelli2021}, these place the allowed abundance in the approximate range $f_{\rm PBH}\lesssim10^{-3}$ (conservative) to $\lesssim10^{-9}$ (aggressive), with no single value a hard exclusion. The minimal seeding requirement, one seed per high-$z$ quasar, $f_{\rm PBH} \sim 10^{-14}-10^{-13}$ across the seed window, sits comfortably below even the most aggressive of these bounds. The whole population scenario ($f_{\rm PBH} \sim 10^{-7}$) is more demanding and lies at or above aggressive accretion bounds ($f_{\rm PBH} \lesssim 10^{-9}$), so it is permitted only under the more conservative limit estimates. Our reference level $\beta\sim10^{-15}$ maps through Eq.~\eqref{eq:beta_to_f} to $f_{\rm PBH}\simeq7\times10^{-10}$, $2\times10^{-10}$, and $7\times10^{-11}$ at $10^{5}$, $10^{6}$, and $10^{7}\,M_\odot$, placing it between the requirement and the limits and between the $f_{\rm PBH}=10^{-8}$ and $10^{-12}$ contours of Fig.~\ref{fig:param_space_scans}. The seed-relevant parameter space is therefore consistent with current constraints. The high-abundance corners are not needed for seeding and are disfavored by the tightest limits.

\section{Discussion and conclusions}
\label{sec:conclusion}

The early assembly of SMBHs places a timing demand on seed formation that is difficult to meet with stellar-mass remnants, and the seed-mass range $10^{5}$--$10^{7}\,M_\odot$ is a natural target for a primordial origin. We have examined whether PBHs can form these seeds under the CMB $\mu$-distortion bound, which limits the small-scale curvature variance over the wavenumbers that collapse to these masses.

Under Gaussian statistics the FIRAS bound, $\sigma_\zeta^{2}\lesssim10^{-4}$, suppresses the seed-mass PBH abundance far below any useful level. This suppression is a property of the Gaussian PDF, in which the variance fixes the tail; it is not a constraint on the tail itself. Because the distortion is a second-moment observable while collapse is a deep-tail observable, a non-Gaussian distribution decouples the two, and the problem reduces to the asymptotic shape of the one-point PDF tens of standard deviations from its center.

The fluctuations that collapse lie well beyond the reach of an $f_{\rm NL}$ expansion, and higher cumulants cannot fatten the tail without disturbing the variance. Working with the exact $\delta N$ map, we organized the possibilities into a compact dictionary and studied four Gaussian-cored families spanning it. The ordinary exponential tail ($D_\infty=k$, $p=1$) generic to single-field non-attractor dynamics, ultra-slow-roll and the upward step, is, despite being far heavier than Gaussian, still too light to reopen the seed window once the distortion bound is imposed, and accounting for the finite width of the step only steepens it further~\cite{Kawaguchi2023} (Sec.~\ref{subsec:results_phys}). The window reopens only for tails heavy enough that $D_\infty=0$: the algebraic tail from fractional-potential dynamics, and the heavier log-normal tail from multiplicative dynamics. The scope of this negative result should be kept in mind: it is the tails realized by the standard non-attractor constructions that fall short; we do not claim that single-field inflation cannot produce sub-exponential tails at all. Our asymmetric, Gaussian-cored families also behave differently from the symmetric global tails analyzed previously~\cite{Hooper2024}, and we map their viable regions directly.

Several open questions remain. The fractional-potential realization of the algebraic tail calls for an explicit construction, whether in non-canonical (DBI) and sound-speed-resonance frameworks~\cite{SilversteinTong2004,Dirac_Born_Infled_SSR} or in multifield and curvaton settings~\cite{LythUngarelliWands2003,AndoInomata2018curvaton}; the log-normal case, which we treat as provisional, calls for a first-principles derivation of the multiplicative dynamics that would generate it. On the formation side, the narrow-feature distortion formula should be replaced by the integral over the actual model spectrum, including the non-Gaussian corrections to the dissipated energy quantified in Refs.~\cite{Sharma:2024img,Byrnes:2024vjt}, and the Press--Schechter criterion by a treatment based on the compaction function and critical collapse, for which the nonlinear relation between $\zeta$ and the density contrast is essential and itself a source of non-Gaussianity~\cite{NiemeyerJedamzik1998,ShibataSasaki1999,MuscoMillerRezzolla2005,Musco2019,YooHaradaGarrigaKohri2018,YoungMuscoByrnes2019,EscrivaGermaniSheth2020,YangCompaction2024}. Finally, the same scales that source these PBHs source a stochastic gravitational-wave background at second order~\cite{SaitoYokoyama2009,KohriTerada2018,Domenech2021,UnalKovetzPatil2021}, providing an independent handle, and the proposed PIXIE mission~\cite{kogut2025primordialinflationexplorerpixie}, with projected sensitivity $\sigma(\mu)\sim10^{-8}$, would tighten the distortion bound by about three orders of magnitude~\cite{Chluba2021}; a non-detection would push the required tail weight further into the heavy regime and correspondingly narrow the class of viable early-universe dynamics.

In summary, PBH formation, and PBH-based constraints on inflation, should be studied in terms of the full non-perturbative one-point PDF rather than the power spectrum or a perturbative non-Gaussianity parameter alone. On this footing the seed-mass window reopens, and a primordial origin for the supermassive black holes observed at high redshift remains viable.

\newpage 

\emph{Note added.} While this work was being finalized, an independent and concurrent study~\cite{Allegrini:2026jqt} appeared that addresses the same problem, non-Gaussian evasion of the CMB $\mu$-distortion bound on supermassive primordial black hole seeds, through a specific derived mechanism, the self-interacting curvaton. Our approach is complementary: rather than a single model, we classify the admissible non-Gaussian tail shapes and identify which of them can reopen the seed window.

\vspace{1 cm}

\begin{acknowledgments}
We thank Junyue Yang, Mian Zhu, Xiao-Han Ma and Swagat Saurav Mishra for valuable discussions. This work was supported in part by the National Key R\&D Program of China (2021YFC2203100), by NSFC (12433002, W2533006), by CAS young interdisciplinary innovation team (JCTD-2022-20), by the 111 Project (B23042), by CSC Innovation Talent Funds, by USTC Funds of International Cooperation, and by USTC Research Funds of the Double First-Class Initiative. SD is supported by USTC with \emph{CAS-ANSO} Scholarship.
\end{acknowledgments}

\bibliography{Refs}
\end{document}